\documentclass[fleqn,usenatbib]{mnras}
\usepackage{newtxtext,newtxmath}

\usepackage[T1]{fontenc}

\DeclareRobustCommand{\VAN}[3]{#2}
\let\VANthebibliography\thebibliography
\def\thebibliography{\DeclareRobustCommand{\VAN}[3]{##3}\VANthebibliography}

\usepackage{graphicx}	
\usepackage{amsmath}	

\title[Motivation for flexible SFHs]{The motivation for flexible star-formation histories from spatially resolved scales within galaxies}

\author[Jain, Tacchella \& Mosleh]{
Shweta Jain,$^{1}$\thanks{E-mail: sja281@uky.edu}
Sandro Tacchella,$^{2,3}$
Moein Mosleh$^{4,5}$
\\
$^{1}$Department of Physics and Astronomy, University of Kentucky, KY, USA\\
$^{2}$Kavli Institute for Cosmology, University of Cambridge, Madingley Road, Cambridge, CB3 0HA, UK\\
$^{3}$Cavendish Laboratory, University of Cambridge, 19 JJ Thomson Avenue, Cambridge, CB3 0HE, UK\\
$^{4}$Biruni Observatory, School of Science, Shiraz University, Shiraz 71946-84795, Iran\\
$^{5}$Department of Physics, School of Science, Shiraz University, Shiraz 71946-84795, Iran}

\date{Accepted 2023 October 27. Received 2023 October 25; in original form 2023 July 27}

\pubyear{2023}

\begin{document}
\label{firstpage}
\pagerange{\pageref{firstpage}--\pageref{lastpage}}
\maketitle

\begin{abstract}
The estimation of galaxy stellar masses depends on the assumed prior of the star-formation history (SFH) and spatial scale of the analysis (spatially resolved versus integrated scales). In this paper, we connect the prescription of the SFH in the Spectral Energy Distribution (SED) fitting to spatially resolved scales ($\sim\mathrm{kpc}$) to shed light on the systematics involved when estimating stellar masses. Specifically, we fit the integrated photometry of $\sim970$ massive (log (M$_{\star}$/M$_{\odot}) = 9.8-11.6$), intermediate redshift ($z=0.5-2.0$) galaxies with \texttt{Prospector}, assuming both exponentially declining tau model and flexible SFHs. We complement these fits with the results of spatially resolved SFH estimates obtained by pixel-by-pixel SED fitting, which assume tau models for individual pixels. These spatially resolved SFHs show a large diversity in shapes, which can largely be accounted for by the flexible SFHs with \texttt{Prospector}. The differences in the stellar masses from those two approaches are overall in good agreement (average difference of $\sim 0.07$ dex). Contrarily, the simpler tau model SFHs typically miss the oldest episode of star formation, leading to an underestimation of the stellar mass by $\sim 0.3$ dex. We further compare the derived global specific star-formation rate (sSFR), the mass-weighted stellar age (t$_{50}$), and the star-formation timescale ($\tau_{\mathrm{SF}}$) obtained from the different SFH approaches. We conclude that the spatially resolved scales within galaxies motivate a flexible SFH on global scales to account for the diversity of SFHs and counteract the effects of outshining of older stellar populations by younger ones.
\end{abstract}



\section{ Introduction } \label{Intro}
The formation of stars from gas and dust is a fundamental process that significantly impacts the development of galaxies throughout the universe. Various factors influence the star-formation (SF) activity \citep[e.g.][]{Iyer_2020, Tacchella_2020}, including sporadic bursts resulting from major mergers \citep{Wellons_2015}, interaction with other galaxies \citep[e.g.][]{Bekki_2011}, episodes of violent disk instabilities and compaction \citep{Dekel_2013, Lapiner_2023}, stellar feedback \citep{Semenov_2021, dome2023miniquenching}, active-galactic nuclei (AGN) feedback mechanisms \citep{Di_Matteo_2005, 2006MNRAS.370..645B, Bluck_2014, Henden_2018}, and dynamical process related to spiral arms and bars \citep{Elmegreen_2011, Shin_2023}. 
These complex processes, in turn, shape the characteristic features of the galaxies, such as the morphological features \citep[e.g.][]{sales2010feedback,dutton2009impact, Tacchella_2016,cochrane2023impact}, the interstellar medium content \citep[e.g.][]{Semenov_2017, Semenov_2021}, and the metallicity content within the galaxies \citep[e.g.][]{Lilly_2013, Forbes_2014}. The detailed study of these SFHs can offer us crucial insights into the processes of galaxy formation and the galaxy's evolutionary pathways over cosmic time.

One technique commonly used to measure the SFHs is modelling and fitting the observed SED of galaxies \citep{Sawicki_1998,  Conroy_2013}. This involves comparing the observed light from a galaxy across a range of wavelengths to models of galaxy evolution that predict how a galaxy's SED changes as it forms stars over time. To generate these SED fitting models, we require a set of priors to infer the galaxy's physical properties (stellar mass, specific star-formation rate [sSFRs], stellar ages t$_{50}$, and attenuation) from the data. These fitting models typically incorporate SFH as a critical component, making the quality of the fitted SFH crucial in determining the reliability and precision of the derived physical properties of the galaxy \citep{2012A&A...541A..85M, 2014A&A...571A..75M,Carnall_2019, Leja_2019}. 

Various approaches can be adopted by SED fitting method for determining the SFHs of galaxies. One such approach is to employ parametric functional forms for the SFHs, such as the exponentially declining tau model, delayed tau model, or lognormal model \citep{Papovich_2010, Gladders_2013, Diemer_2017}, and parametric models with additional flexibilities incorporating multiple episodes of SF \citep{Morishita_2019, Ciesla_2017, French_2018}. The restricted nature of these parameterised forms with a fixed number of parameters are unlikely to capture the rich diversity of galaxy SFHs.
As a result, the inferred galaxy properties based on these SFHs may be subject to systematic biases, with potential under-reporting of uncertainties, see, e.g., \citet{Walcher_2010,Conroy_2013,https://doi.org/10.48550/arxiv.1404.0402,  Diemer_2017,https://doi.org/10.48550/arxiv.2212.01915,Carnall_2019,Whitler_2023}. A solution to these issues is to use flexible or non-parametric models that do not assume any explicit functional form and allow for arbitrary SFRs as a function of time, allowing to capture the complexity of physical SFHs. Examples include piecewise constant SFRs in time \citep{Ocvirk_2006, Leja_2017, Leja_2019, Morishita_2019}, the dense Basis SFH reconstruction method \citep{Iyer_2017, Iyer_2019}, and libraries of SFHs measured from theoretical models of galaxy formation \citep{Pacifici_2012}. 

Though flexible SFH models are more computationally expensive than parametric ones, they promise more reliable recovered SFHs. For instance, \citet{Lower_2020} tested the galaxy properties inferred from the non-parametric model using the SED fitting code \texttt{PROSPECTOR} \citep{Johnson_2021} by ground-truthing them against mock observations. On the other hand, \citet{Leja_2019} tested the robustness of the inferred galaxy properties by comparing them with that inferred from the parametric models. The conclusion from these works is that though the flexibility offered by the non-parametric priors is a clear advantage of this approach, yielding robust and reliable results, there remains a potential weakness in the dependence of the SFHs on the prior selection \citep{Suess_2022, Tacchella_2022,2023MNRAS.519.5859W}.

A complimentary view on the issue of SFH model and prior selection can be gained by using spatially resolved information from galaxies, i.e., by extracting SFHs from spatially resolved scales \citep[e.g.,][]{2018MNRAS.476.1705S}. Most of the findings described above rely on either single-aperture spectroscopic data or integrated multi-band photometric data. By studying the galaxies using integrated photometry, we can only study the statistical evolution of the stellar populations in galaxies as a whole, but it is not easy to understand SF activity happening on small scales which actually trace the galaxy evolution pathways. Additionally, several works have reported that the spatial resolution of the data can also introduce significant biases in the stellar mass estimates and other inferred properties of the galaxies \citep{Zibetti_2009, Wuyts_2012, Sorba_2015, Suess_2019, Mosleh_2020, Abdurro_uf_2022}. With a similar motivation, \citet{Mosleh_2020} studied the spatial distribution of stellar mass in a galaxy using the spatially resolved (i.e., pixel-by-pixel) broad-band SED fitting tool, introduced by \citet{Abraham_1999} and \citet{Conti_2003}, and found a clear bias between the stellar masses estimated on unresolved and resolved scales. This study motivated us to investigate the impact of spatial resolution on the recovered SFHs, and how this affects the determination of physical properties such as stellar masses, sSFRs, t$_{50}$, and $\tau_{\mathrm{SF}}$.

Being able to estimate accurate SFHs is fundamentally important not only for galaxies in the local Universe, but also to interpret the light emission of early galaxies. JWST is revolutionizing how we see high-redshift galaxies, though deriving physical quantities from those observations is challenging because galaxies' SFHs are expected to be more variable (i.e. burstier; \citealt{Faucher_Gigu_re_2017,Tacchella_2020}). This leads to an increased importance of outshining, where the most recent burst of star formation dominates the SED, making it difficult to assess the stellar mass in the older stellar populations \citep[e.g.,][]{narayanan2023outshining}. Early JWST results indeed indicate that many of the rest-UV bright galaxies are undergoing bursts of star formation \citep{endsley2023starforming,Laporte_2023,looser2023jades,Tacchella_2023}, though accreting black holes are also contributing and can lead to difficulty to interpret the SEDs \citep{Kocevski_2023,larson2023ceers,maiolino2023small,matthee2023little,harikane2023jwstnirspec}. One way to address these challenges is to spatially resolve these early galaxies and connect the global SED to the resolved SEDs \citep[See for first results:][]{https://doi.org/10.48550/arxiv.2212.08670,Abdurro_uf_2023,2023arXiv230602472B,2023ApJ...952...74T,2023ApJ...949L..34H}.

In this study, we present detailed measurements of SFHs both on global and spatially-resolved scales for a sample of $\sim$970 distant galaxies with redshifts $z = 0.5-2.0$ to shed light on the systematics involved when estimating galaxy properties and to motivate more complex SFHs on global scales. We adopted four types of simple and flexible models for the SFHs, both on spatially resolved and unresolved scales. 
On spatially resolved scales, we derive SFHs of individual pixels using pixel-by-pixel SED fitting adopting iSEDfit, which we then combine to a total SFHs by summing the pixel-based SFHs (SFH$_{\rm \star, res}$). On global scales, we fit the integrated photometry using (i) a simple tau model within iSEDfit (SFH$_{\rm \star, int,\tau}$), (ii) a simple tau model within \texttt{Prospector} (SFH$_{\rm \star, int,non-flex}$), and (iii) a flexible, non-parametric model within \texttt{Prospector} (SFH$_{\rm \star, int,flex}$).

The organisation of the paper is as follows. In Section \ref{Method}, we briefly overview the data sets and methodology adopted to get the SFHs of the galaxies from different models. In Section \ref{Results}, we determine the reconstructed SFHs of the galaxies with all the models and compare the inferred galaxy properties from these SFHs. The core of this paper is Section \ref{Dis}, where we discuss the implications of how the assumed SFH models and the spatial resolution affect the inferred physical properties of the galaxies. We summarise our results in Section \ref{Conc}. 

\begin{figure}
\centering
\includegraphics[width=0.5\textwidth]{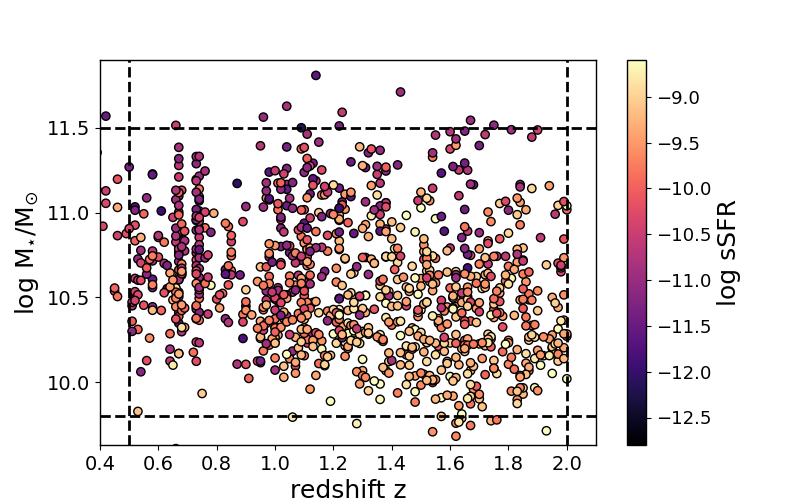}
\caption{The stellar mass-redshift distribution of the sample of galaxies colour-coded by sSFR, where the SFR is measured over the last 100 Myr. The fiducial model is used to infer sSFRs and stellar masses. The photometric redshifts ($z$) of galaxies are determined using EAZY code \citep{Brammer_2008} and taken from 3D-HST catalogue (\citet{2014ApJS..214...24S}; See \citet{Mosleh_2020}). The stellar mass and redshift limits of the sample are $9.8 \leq \log$(M$_{\star}$/M$_{\odot}) \leq 11.5$ and $0.5 \leq z \leq 2.0$, respectively. These cuts on the stellar masses and the redshifts are shown by the dashed lines, giving us a sample of 967 galaxies.} 
\label{M-Z_range}
\end{figure}

\begin{figure*}
\includegraphics[width=\textwidth]{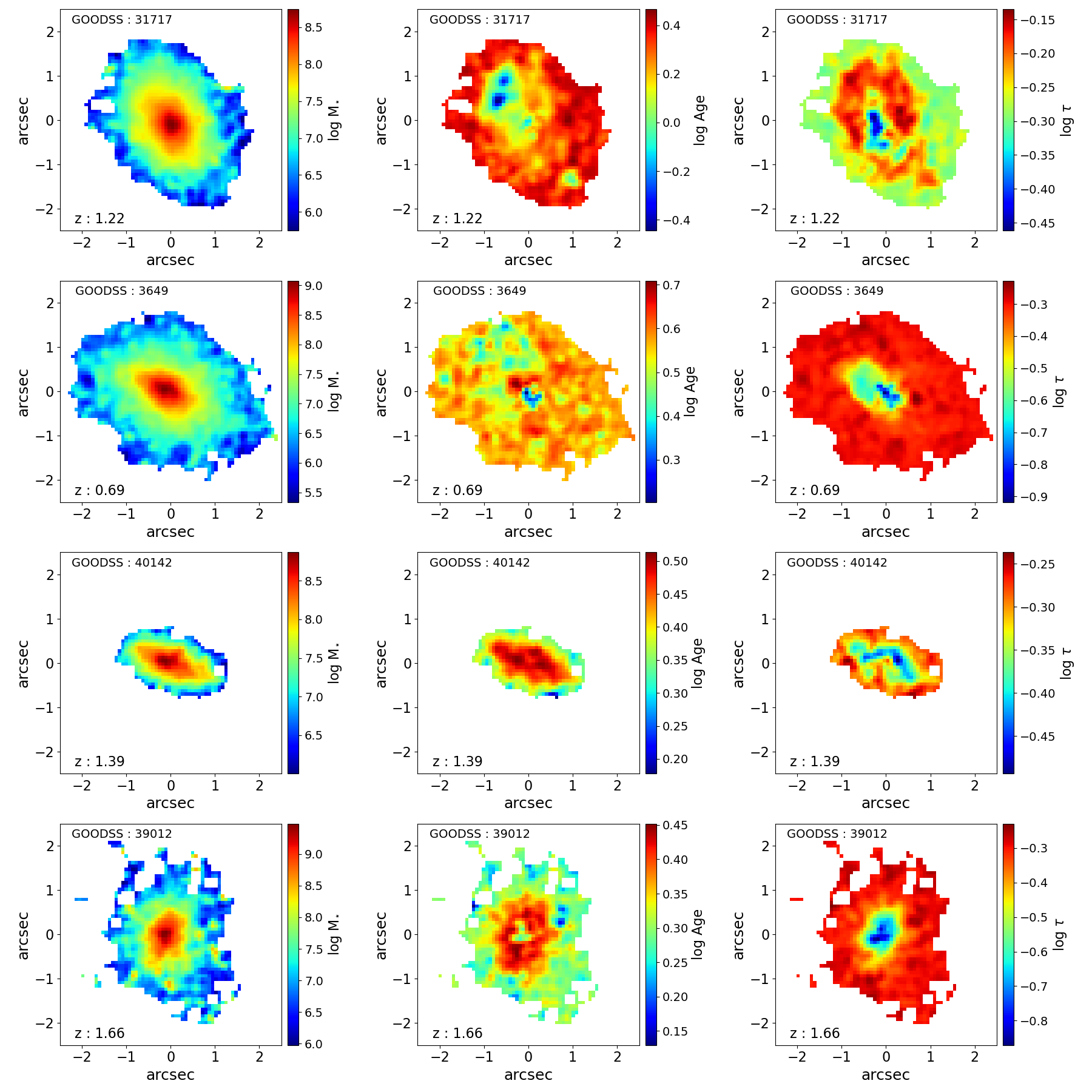}
\caption{The left, middle, and right panels show the inferred stellar mass maps (in $M_{\odot}$), stellar age maps (in Gyr), and e-folding time ($\tau$) maps (in Gyr) of four example galaxies. These maps are derived using the pixel-by-pixel SED fitting method \citep{Mosleh_2020}. The maps obtained are smoothed by a Gaussian kernel of $\sigma=1$ pixel.}
\label{F_maps}
\end{figure*}

\begin{figure*}
\centering
\includegraphics[width=\textwidth]{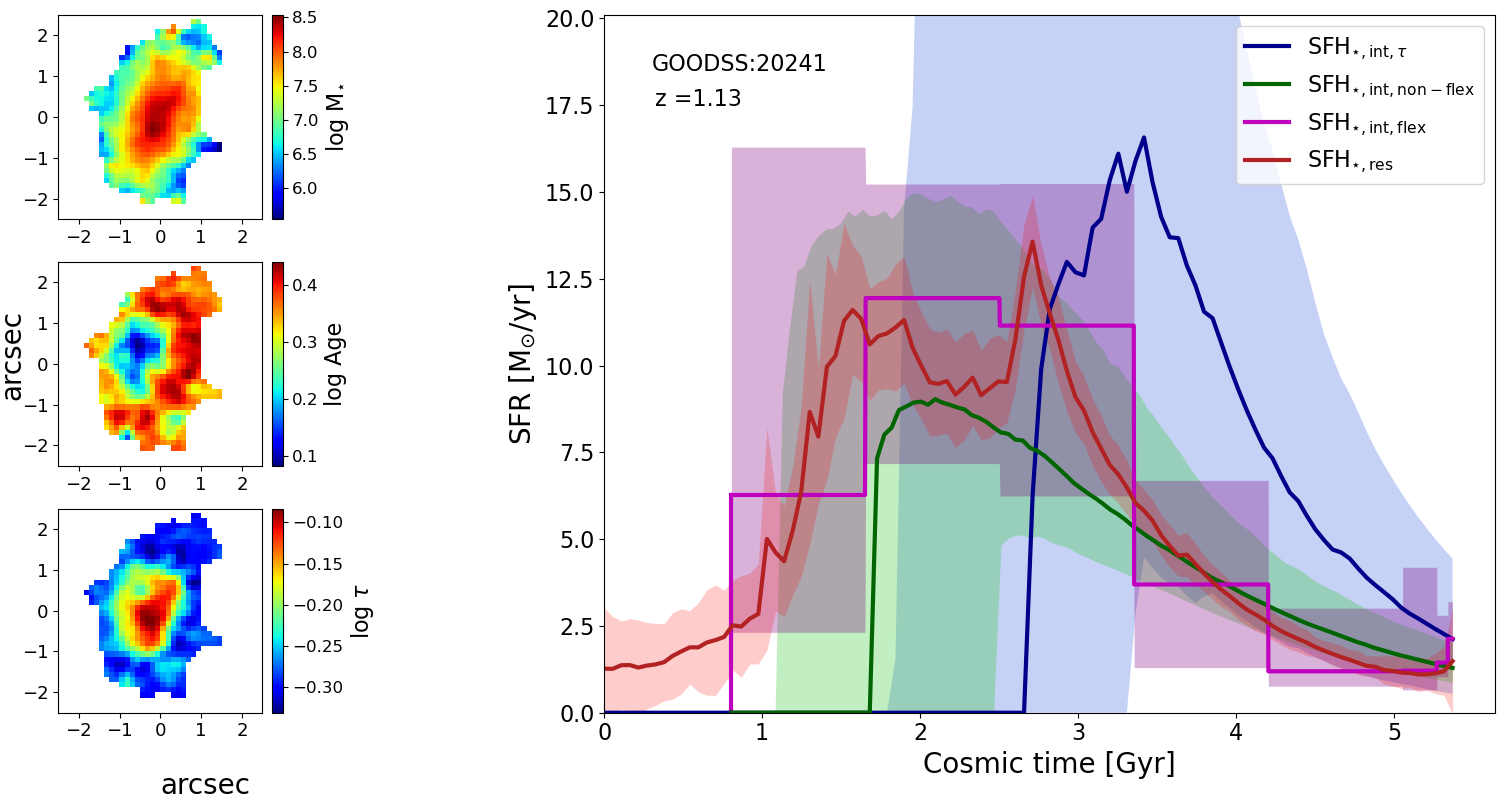}
\caption{SFH of an example galaxy. The left panels show the stellar mass map, age map, and $\tau$ map in the first, second, and third rows, respectively. 
The right panel shows the four SFHs: (a) the global average spatially resolved SFH obtained from pixel-by-pixel SED fitting (red line; SFH$_{\rm \star, res}$); (b) the SFH derived from the SED fitting of the total fluxes of all the pixels using iSEDfit code (blue line; SFH$_{\rm \star, int,\tau}$); (c) the SFH derived from the \texttt{Prospector} model that assumes simple tau-model (green line; SFH$_{\rm \star, int,non-flex}$); and (d) the SFHs obtained from \texttt{Prospector}, which adopts a flexible SFH (purple line; SFH$_{\rm \star, int,flex}$). The shaded regions indicate the 16$^{\mathrm{th}}$-84$^{\mathrm{th}}$ percentile range in each case (see Sections \ref{R_SFHs} and \ref{Prospector} for details).
The spatially resolved SFH represents a variable SFH expected from a physical SFH. The fiducial, \texttt{Prospector}-based SFH is able to trace the overall shape well. The SFHs from SFH$_{\rm \star, int,\tau}$ and SFH$_{\rm \star, int,non-flex}$ not able to capture this shape and is typically biased young, i.e. misses older stellar populations.}
\label{SFH_resolved}
\end{figure*}

\begin{figure*}
\includegraphics[width=\textwidth]{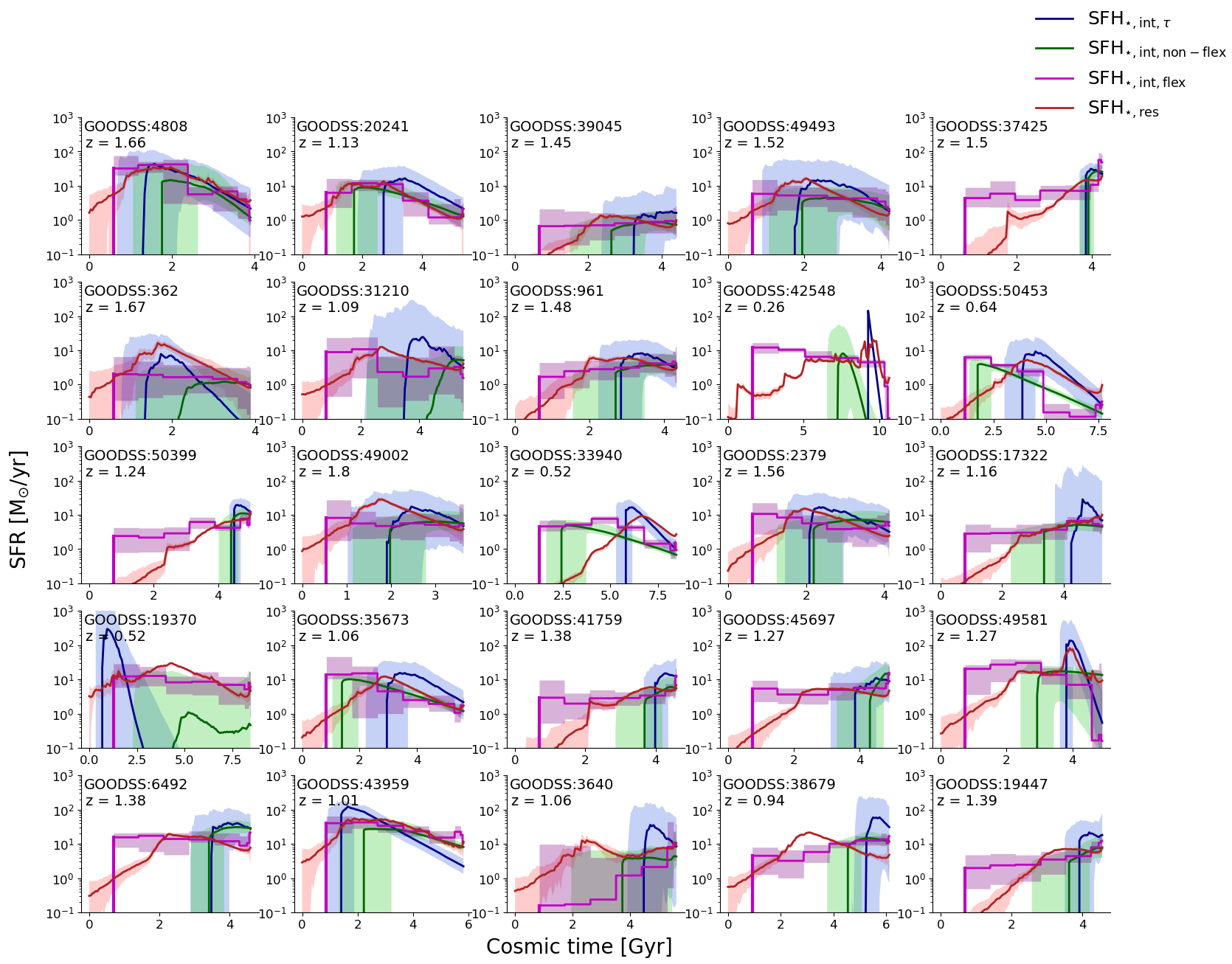}
\caption{Comparing the SFHs for the four models in this paper (Sections \ref{R_SFHs} and \ref{Prospector}). The four SFHs are: (a) The spatially resolved SFH obtained from pixel-by-pixel SED fitting (red curve; SFH$_{\rm \star, res}$); (b) The SFH derived from the SED fitting of the total fluxes of all the pixels using iSEDfit code (blue curve; SFH$_{\rm \star, int,\tau}$); (c) The SFH derived from the \texttt{Prospector} model that assumes simple tau-model (green curve; SFH$_{\rm \star, int,non-flex}$); and (d) The SFHs obtained from \texttt{Prospector}, which adopts a flexible SFH as our fiducial model in this paper (purple curve; SFH$_{\rm \star, int,flex}$). The shaded regions indicate the 16$^{\mathrm{th}}$-84$^{\mathrm{th}}$ percentile range in each case (See Section \ref{R_SFHs} and \ref{Prospector} for details). The spatially resolved SFH covers a vast age range and represents a more variable SFH similar to the fiducial SFH (SFH$_{\rm \star, int,flex}$). The SFHs from SFH$_{\rm \star, int,\tau}$ and SFH$_{\rm \star, int,non-flex}$ not able to capture this diversity and is typically biased young, i.e. misses older stellar populations.}
\label{SFHs}
\end{figure*}

\section{Methods} \label{Method}
\subsection{ Data }

In order to perform pixel-by-pixel SED fitting \citep{Abraham_1999, 
Conti_2003, Zibetti_2009}, the measurements of the stellar mass, age,
and $\tau$ distributions as well as SFHs on kpc-scale for $\sim$ 970 
galaxies have been taken from \citet{Mosleh_2020}. The galaxy sample has been confined to a redshift range of $0.5 \leq z \leq 2.0$ and a stellar mass range of $9.8 \leq$ log (M$_{\star}$/M$_{\odot})$  $\leq 11.5$. The lower redshift bound and the upper stellar mass bound are motivated by completeness due to volume effects. The lower stellar mass bound is motivated by completeness, because of detecting faint galaxies. The upper redshift limit ensures that the galaxies' rest-frame optical SEDs are probed by several filters.

\autoref{M-Z_range} plots the ranges of stellar masses and redshifts of the galaxies used for the sample. 
The author utilised the publicly available catalogue and imaging dataset from the 3D-HST Treasury 
Programme \citep{2012ApJS..200...13B,2014ApJS..214...24S} and the 
Cosmic Assembly Near-IR Deep Extragalactic Legacy Survey (CANDELS; \citealt{2011ApJS..197...35G,2011ApJS..197...36K}) to conduct their 
study. 
We use GOODS-South field data because of the largest number of filters and best depths, i.e., the highest S/N ratio on spatially resolved scales for individual galaxies \citep{Guo_2013}. We utilise PSF-matched mosaic images of GOOD-South field data, comprising a maximum of seven filters ($B_{435}, V_{606}, i_{775}, z_{850}, J_{125}, JH_{140}, H_{160}$).
The total area of this field is about $170$ arcmin$^2$. Using these ancillary data, the stellar masses and photometric redshifts (if no spectroscopic or grism redshift is available) determined with the EAZY \citep{2008ApJ...686.1503B} and FAST \citep{2009ApJ...700..221K}
codes, respectively.  
\subsection{ Creating 2d Stellar Maps }\label{2D}
With the setup described in \citet{Mosleh_2020}, the spatially resolved SED fitting (pixel-by-pixel method) method is used to obtain the spatially resolved physical maps such as mass maps, age maps, and $\tau$ maps of each of the galaxies. The pixels have a size of 0.06 arcsec, corresponding to 0.38 and 0.52 kpc at $z=0.5$ and $z=2.0$, respectively. The best-fit SED model for each pixel is used to obtain these resolved stellar properties' maps. They used iSEDfit \citep{Moustakas_2013}, a Bayesian code, to perform the SED fitting. They created a full grid of 100,000 models based on the \citet{2003MNRAS.344.1000B} stellar population evolution models with ages between 0.1 and 13.5 Gyr.
The SFHs for these models are assumed to be exponentially declining ($\hbox{SFR} \propto$ exp(-t$/\tau)$), with the e-folding timescale $\tau$ between (0.01 - 1.0 Gyr) and the \citet{2003PASP..115..763C} initial mass function (IMF) is adopted. This exponentially declining model is also known as simple tau model. The metallicity range used is 0.004-0.03, and the \citet{2000ApJ...533..682C} dust attenuation law is assumed. For each galaxy, the redshift of all pixels used is the redshift of the galaxy from the 3D-HST catalogue. To study the spatial distribution of the physical properties of the galaxies, \autoref{F_maps} shows the mass, age, and $\tau$ maps of a few galaxies in the left, middle, and right panels, respectively. It displays the 50$^{\mathrm{th}}$ percentile of the inferred parameters.

\subsection{SFHs from spatially resolved scales}\label{R_SFHs}

Using the derived maps of stellar mass, age, and $\tau$ from \citet{Mosleh_2020}, we estimate first the SFH of each individual pixel and then combine those to obtain the total SFH (SFH$_{\rm \star, res}$), ensuring propagation of the errors. For a galaxy observed when the age of the universe was $t_{\text{obs}}$ Gyr old, the SFH assuming a simple tau model can be calculated as follows:
\begin{align}
\text{SFR}(t) &=
\begin{cases} 
   \mathrm{norm} \cdot e^{-\frac{(t - (t_{\text{obs}}-t_{\text{age}}))}{\tau}} & \text{if } t > t_{\text{obs}} - t_{\text{age}}\\
   0 & \text{if } t < t_{\text{obs}} - t_{\text{age}}
\end{cases}
\label{eq1}
\end{align}
with normalisation factor $\mathrm{norm}$ given by,
\begin{equation}
    \mathrm{norm} = \frac{ M_{\star}}{\int_{0}^{t_{\text{obs}}} e^{-\frac{(t - (t_{\text{obs}}-t_{\text{age}}))}{\tau}} dt}
\end{equation}
where  $M_{\star}$ is the total stellar mass, $t_{\text{age}}$ is the lookback age when SFH started, and $\tau$ represents the e-folding timescale. 
For each galaxy, we fix the stellar mass associated with each $j^{\mathrm{th}}$ pixel and draw $N$ values of both $t_{\text{age}}$ and $\tau$ from the Gaussian distribution (with 1 $\sigma$ uncertainty). We take $N = 100$. 
This results in two $N$ draws: the first set consider the variation in age, keeping stellar mass and $\tau$ fixed to their best-fit values, \begin{equation}
    \left\{\left(M_{\star, j}, t_{\text{age},j}^i, \tau_j\right)\right\},\quad \mathrm{and}
\end{equation}
the second set considers the variation in $\tau$, keeping stellar mass and age fixed,
\begin{equation}
    \left\{\left(M_{\star, j}, t_{\text{age},j}, \tau_j^i\right)\right\} \quad \forall \, i \in [1,N=100].
\end{equation}
Using the above parameter sets and assuming a simple tau model (see Equation \ref{eq1}), results in two distinct sets of $N$ SFHs for each pixel: one accounting for errors in stellar age $t_\mathrm{age}$, \{$\mathrm{SFH}_{j}(M_{\star, j}, t_{\text{age},j}^i, \tau_j$)\} and another accounting for errors in $\tau$, \{$\mathrm{SFH}_{j} (M_{\star, j}, t_{\text{age},j}, \tau_j^i)$\}. 
Next, we sum the SFHs 
\{$\mathrm{SFH}_j$\} associated with same varied parameter ($\tau$ or $t_{\text{age}}$) across all pixels , resulting in two sets of $N$ global SFHs for the entire galaxy:
\begin{equation}
    \{\mathrm{SFH}_{i}(t)\}_{\tau} = \left\{\sum_{j=1}^{\mathrm{No. \, of \, pixels}} \mathrm{SFH}_{j}\left(M_{\star, j}, t_{\text{age},j}, \tau_j^i\right)\right\}, \quad \text{and}
\end{equation}
\begin{equation}
    \{\mathrm{SFH}_{i}(t)\}_{t_{\text{age}}} = \left\{\sum_{j=1}^{\mathrm{No. \, of \, pixels}} \mathrm{SFH}_{j}\left(M_{\star, j}, t_{\text{age},j}^i, \tau_j\right)\right\} \; \text{for} \; i \in [1,N=100]
\end{equation}
where the subscript outside the curly bracket on LHS represents the parameter varied to obtain the SFHs set.
Then, to obtain resolved SFHs that accounts for the error in parameter-$\tau$ ($\mathrm{SFH}_{\star, \mathrm{res}, \tau}(t)$), and $t_\mathrm{age}$ ($\mathrm{SFH}_{\star, \mathrm{res}, t_{\text{age}}}(t)$), we take the median of the respective sets containing $N$ SFHs,
\begin{equation}
    \mathrm{SFH}_{\star, \mathrm{res}, \tau}(t) = \text{median}\left(\{\mathrm{SFH}_{i} (t)\}_{\tau}\right), \text{ and }
\end{equation}
\begin{equation}
    \mathrm{SFH}_{\star, \mathrm{res}, t_{\text{age}}}(t) = \text{median}\left(\{\mathrm{SFH}_{i} (t)\}_{ t_{\text{age}}}\right) \; \forall \; i \in [1,N=100]
\end{equation}
Finally, we calculate the global SFH ($\mathrm{SFH}_{\star, \mathrm{res}} (t)$) by averaging the resolved SFH accounting for the error in $\tau$ and $t_\mathrm{age}$,
\begin{equation}
    \mathrm{SFH}_{\star, \mathrm{res}} (t)= \frac{\mathrm{SFH}_{\star, \mathrm{res}, \tau}(t)+\mathrm{SFH}_{\star, \mathrm{res}, t_\mathrm{age}}(t)}{2}.
\end{equation}

\autoref{SFH_resolved} is a schematic figure illustrating the obtained global SFH from the pixel-by-pixel information. The left panels of \autoref{SFH_resolved} plot the stellar mass, age, and $\tau$ map of an example galaxy. The smooth maps highlight the spatially varying galaxy parameters used to calculate SFHs. The right panel shows the global SFH inferred from the spatially resolved maps (we call this the spatially resolved SFH; SFH$_{\rm \star, res}$; red line), the SFH derived from the SED fitting of the total fluxes of all the pixels using iSEDfit code (blue line; SFH$_{\rm \star, int,\tau}$), the SFH derived from the \texttt{Prospector} model that assumes simple tau-model (green line; SFH$_{\rm \star, int,non-flex}$), and the SFHs obtained from \texttt{Prospector}, which adopts a flexible SFH as our fiducial model in this paper (purple line; SFH$_{\rm \star, int,flex}$). The shaded regions indicate the 16$^{\mathrm{th}}$-84$^{\mathrm{th}}$ percentile range in each case.

\subsection{SFH from integrated photometry}\label{Prospector}
We compare the SFHs from spatially resolved scales to the ones from integrated photometry. We obtain the integrated photometry by summing the fluxes of all pixels that belong to the galaxy, as identified in the stellar mass maps in Figure~\ref{F_maps}. This ensures that any differences in the SFHs and stellar masses are not based on aperture effects. We then derive SFHs and stellar masses from the integrated photometry in three different ways, assuming: (i) a simple tau model within iSEDfit (SFH$_{\rm \star, int,\tau}$), (ii) a simple tau model within \texttt{Prospector} (SFH$_{\rm \star, int,non-flex}$), and (iii) a flexible, non-parametric model within \texttt{Prospector} (SFH$_{\rm \star, int,flex}$). We run iSEDfit (approach (i)) with the same setup as described in Section \ref{2D} and \citet{Mosleh_2020}. We obtain a single value of derived stellar mass, age, $\tau$ for the entire galaxy from this approach. To ensure the propagation of errors while estimating the SFH from these derived parameter values, we extracted $N$ parameter values within 1$\sigma$ uncertainty (similar to Section \ref{R_SFHs}) using Gaussian distributions in $\tau$ and age. Finally, we take the 50$^{\mathrm{th}}$ percentile (Figure \ref{SFHs}; solid blue line) of these SFHs and take the 16$^{\mathrm{th}}$-84$^{\mathrm{th}}$ percentiles corresponding to the shaded region in blue (see Figure \ref{SFHs}). 

For approaches (ii) and (iii), we use the Bayesian inference SED-fitting code \texttt{Prospector} \citep{Johnson_2021}, which adopts the Flexible Stellar Population Synthesis (\texttt{FSPS}) package \citep{Conroy_2009} for stellar population synthesis. In this work, we use the MIST stellar evolutionary tracks and isochrones \citep{Choi_2016, Dotter_2016}. We adopt a similar \texttt{Prospector} model as outlined in \citet{Leja_2017, Leja_2019, Tacchella_2022}. Specifically, the redshift is set fixed to the photometric redshift (or spectroscopic redshift when available). We adopt a single stellar metallicity that is varied with a prior that is uniform in $\log(Z_{\star}/Z_{\odot})$ between $-1.0$ and 0.19, where $Z_{\odot}=0.0142$. We assume a flexible attenuation law, where we tie the strength of the UV dust absorption bump to the best-fit diffuse dust attenuation index, following the results of \citet{Kriek_2013}.
The dust attenuation law index $n$ is a multiplicative factor relative to the \citet{Calzetti_2000} attenuation curve. We parameterise the dust attenuation curve in accordance with the prescription outlined in \citet{2009A&A...507.1793N},
\begin{equation}
    \tau_{\lambda} = \frac{\tau_V}{4.05} \left(k(\lambda)+E_b D(\lambda)\right)\left(\frac{\lambda}{5500 \text{\AA}}\right)^n
\label{eqn_curve}
\end{equation}
with
\begin{equation}
    E_b = 0.85 - 1.9n.
\end{equation}
In the above Equation \ref{eqn_curve}, $k(\lambda)$ is the (fixed) \citet{Calzetti_2000} attenuation curve, $D(\lambda)$ is a Lorentzian-like Drude profile describing the UV dust bump at 2175 \AA \, $E_b$ represents an empirical correlation between the strength of the 2175\AA \, dust absorption bump and the slope of the
curve, and  $\tau_V$ is the optical depth of the diffuse component in the V
band. 
The free parameters in this equation are $\tau_{\lambda}$, which controls the normalization of the diffuse dust, and $n$. We assume flat prior for $n \in (-1, 0.4)$.

We run two different versions of the \texttt{Prospector} model based on different assumptions regarding the SFH. In approach (ii), we assume a simple $\tau$ model that has three free parameters: the total stellar mass (uniform prior in log-space between $10^9$ and $10^{12}~M_{\odot}$), the start of the SFH (flat prior between 0.001 Gyr and $85\%$ of the age of the universe at the galaxy's redshift), and $\tau$ (uniform prior in log-space between 0.01 and 30 Gyr). On the other hand, in approach (iii), we adopt a flexible model, where we assume that the SFH is step-wise constant in 8 time bins. We fit for the ratio of the SFR in those bins (7 free parameters) plus the total stellar mass. We use the standard continuity prior \citep{Leja_2019} and assume a uniform prior for the stellar mass in the range of $10^9$ and $10^{12}~M_{\odot}$. The first three-time bins are fixed to $0-30$, $30-100$, and $100-300$ Myr, while the remaining bins are logarithmically in time up to a lookback time of $85\%$ of the age of the universe at the galaxy's redshift. 

Throughout this work, we assume that the flexible \texttt{Prospector} model is our fiducial model against which we compare the other SFHs. Recent works have highlighted the flexibility of these models compared to parametric models \citep[see for e.g.,][]{Leja_2019,Lower_2020,Suess_2022}. We choose the flexible \texttt{Prospector} model as our fiducial model because it has been widely adopted in the literature \citep{Leja_2017, Leja_2019,Suess_2022,endsley2023starforming,Tacchella_2023, 2023ApJ...952...74T} and integrated photometry is readily available (in comparison with spatially resolved SEDs).
The purpose of choosing a fiducial model is to test how well different SFH models are able to recover the properties of galaxies and agree with each other. It is important to note that the paper is not aimed at ground-truthing the accuracy of various SFH approaches but to test the reliability of different approaches when compared to each other. In principle, we expect the SFH from resolved SED fitting should give the most reliable results, because when analyzing a single pixel, the star-dust geometry is simplified, and there may be less variation in age and metallicity, making the results more trustworthy. However, since many pixels need to be fit, the SED models on spatially resolved scales make usually simplified assumptions (for example a simple tau model for the SFH in our case here) in order to shortening the run-time of the fitting (see Section \ref{R_SFHs}).

\begin{figure*}
\centering
\includegraphics[width=\textwidth]{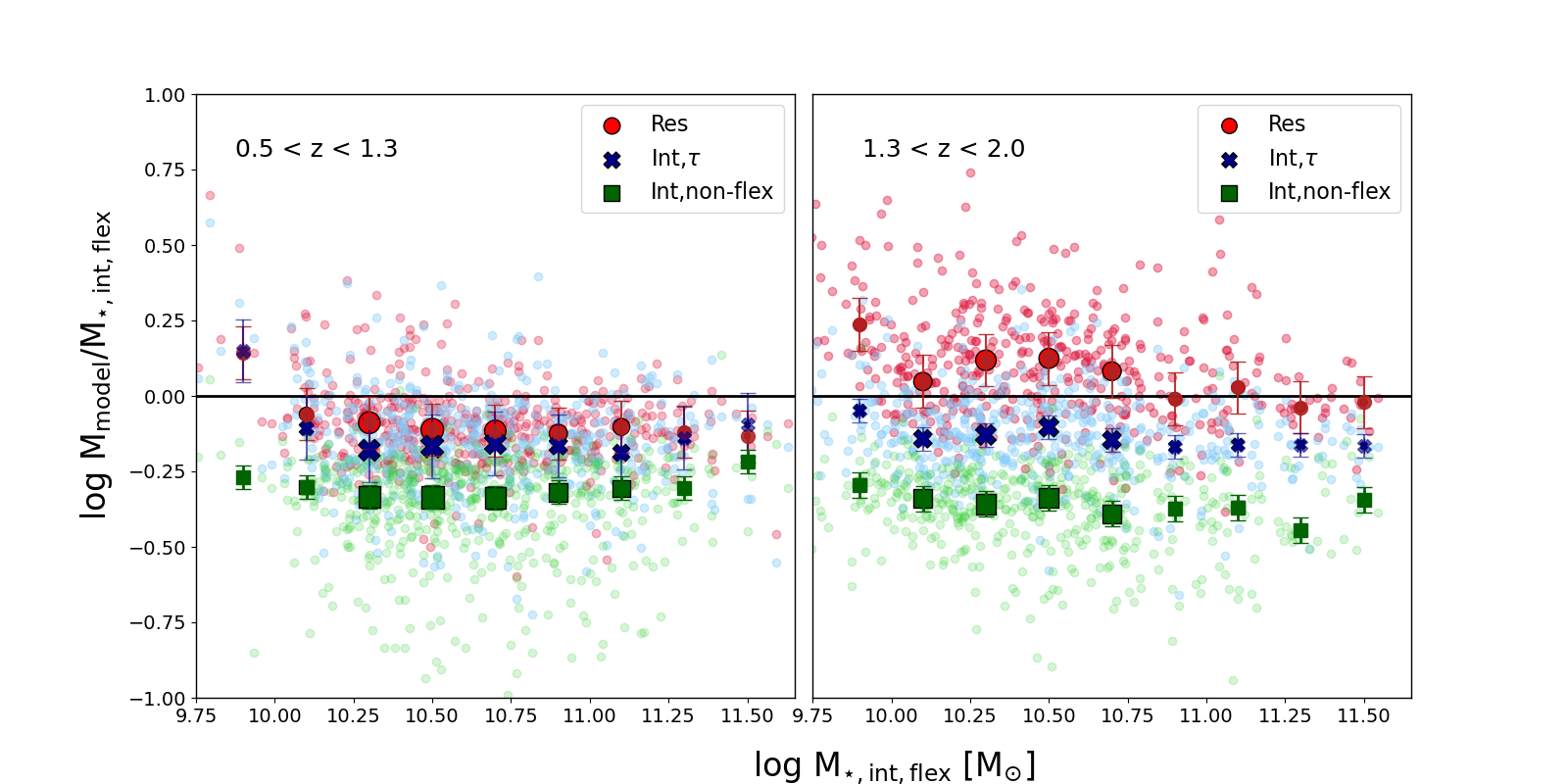}
\caption{Stellar mass differences (in log) for the different SFH models with respect to the fiducial one (flexible \texttt{Prospector} SFH: M$_{\rm \star, int, flex}$) for galaxies with $0.5<z<1.3$ (left panel) and $1.3<z<2.0$ (right panel). The three models include the spatially resolved model from pixel-by-pixel SED fitting (red circles), a simple tau model within iSEDfit (blue crosses), and a simple tau model within \texttt{Prospector} (green squares). The mass differences obtained from the fiducial model and the pixel-by-pixel SED fitting model are the smallest. In contrast, the mass differences obtained from the fiducial model and the \texttt{Prospector} model with a tau-SFH are the largest. 
 Moreover, the stellar masses are underestimated by the SFH models from integrated scales (SFH$_{\rm \star, int,\tau}$ and SFH$_{\rm \star, int,non-flex}$) when compared to the fiducial masses, for both the redshift ranges. This is because those SFHs are unable to capture the early prolonged SF activity from older stellar populations missing out stellar masses due to outshining effects. (see Fig.~\ref{SFHs}).}
\label{Fig_md}
\end{figure*}

\section{ Results } \label{Results}

In this section, we compare and discuss differences between the galaxies' reconstructed SFHs using all the approaches described in Sections \ref{R_SFHs} and \ref{Prospector}. We aim to evaluate the consistency with which we can infer a galaxy's physical properties from the four different approaches. Our analysis demonstrates that spatially resolved SFHs (SFH$_{\star,\mathrm{res}}$) and the flexible \texttt{Prospector} SFH (SFH$_{\rm \star, int,flex}$) are the most consistent with each other in deriving the galaxies' physical properties, while the simpler, parametric SFH approaches are too simplistic to account for the physical diversity of the SFHs. 

\subsection{Overall shapes of galaxy SFHs}\label{Results_SFH}

\autoref{SFHs} shows the recovered SFHs for randomly selected galaxies using the four approaches discussed in Sections \ref{R_SFHs} and \ref{Prospector}. We compare the SFHs obtained from the pixel-by-pixel SED fitting method (SFH$_{\star,\mathrm{res}}$; red line), SED fitting of the total fluxes of all the pixels using iSEDfit code (SFH$_{\star,\mathrm{int},\tau}$; blue line), parametric fitting models from the \texttt{Prospector} modelling (SFH$_{\star,\mathrm{int},\mathrm{non-flex}}$; green line) and the flexible non-parametric fitting models from the \texttt{Prospector} modelling (SFH$_{\star,\mathrm{int},\mathrm{flex}}$; purple line) taken as the fiducial one. The shaded regions indicate the $16^{\mathrm{th}}$ - $84^{\mathrm{th}}$ percentiles. 

For most of the galaxies, the SFH$_{\star,\mathrm{res}}$ are tracing the fiducial SFH$_{\star,\mathrm{int},\mathrm{flex}}$ much better than the SFH$_{\star,\mathrm{int},\tau}$ and SFH$_{\star,\mathrm{int},\mathrm{non-flex}}$. The most prominent feature of SFH$_{\star,\mathrm{res}}$ is that it is able to capture the stochastic behaviour of SF expected from the physical SFH of a galaxy better than the other two models when compared to our fiducial one. 

Furthermore, SFH$_{\star,\mathrm{int},\tau}$ and SFH$_{\star,\mathrm{int},\mathrm{non-flex}}$ fail to match the variations and amplitude of our fiducial SFHs at early cosmic times for most of the galaxies. This may result in missing of a significant portion of the formed stellar mass, potentially leading to an underestimation of the stellar masses of galaxies and other related properties that rely on the assumed shape of the SFHs.

\subsection{Differences in Stellar Masses} \label{R_Diff_mass}

\begin{figure*} 
\includegraphics[width=\textwidth]{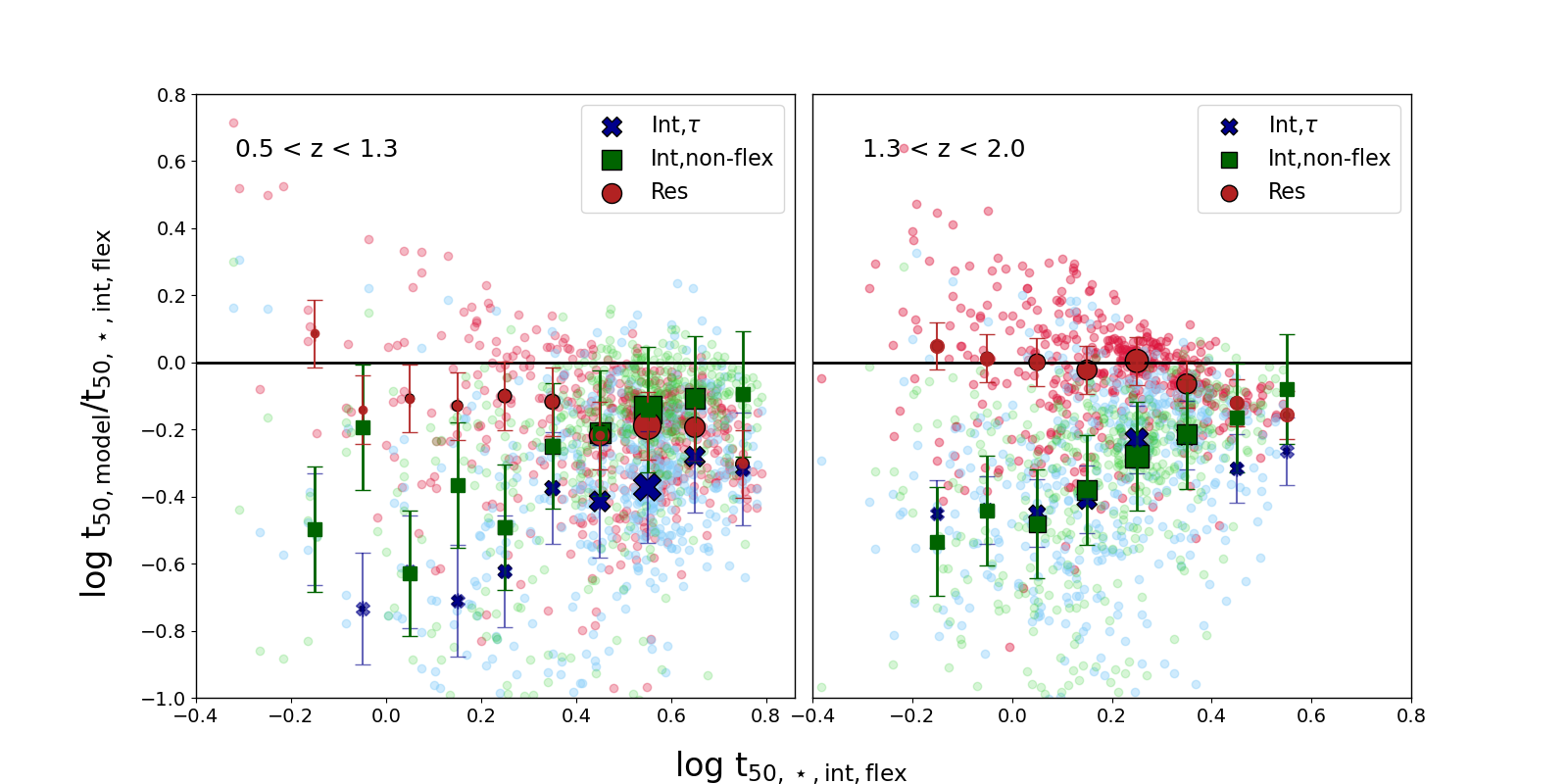}
\caption{The differences in mass-weighted ages (t$_{50}$) inferred from the different SFH models with respect to our fiducial model (flexible \texttt{Prospector} SFH: t$_{50,\rm \star, int, flex}$) for galaxies with $0.5<z<1.3$ and $1.3 < z < 2.0$ (left and right panels, respectively). The figure follows the same layout as Fig.~\ref{Fig_md}. The t$_{50}$ differences obtained from the fiducial and pixel-by-pixel SED fitting models are the least. In contrast, the differences are the largest between the t$_{50}$ obtained from the fiducial model and a simple tau model within iSEDfit from integrated scales. Stellar ages obtained from the SFH models from integrated scales (SFH$_{\rm \star, int,\tau}$ and SFH$_{\rm \star, int,non-flex}$) are much younger (by a factor of 3) when compared to those obtained from the fiducial ages. Moreover, the discrepancy increases for younger galaxies (i.e., galaxies that recently assembled their stellar masses), highlighting the effects of outshining.}
\label{T50_d}
\end{figure*}

\begin{figure*}
\includegraphics[width=\textwidth]{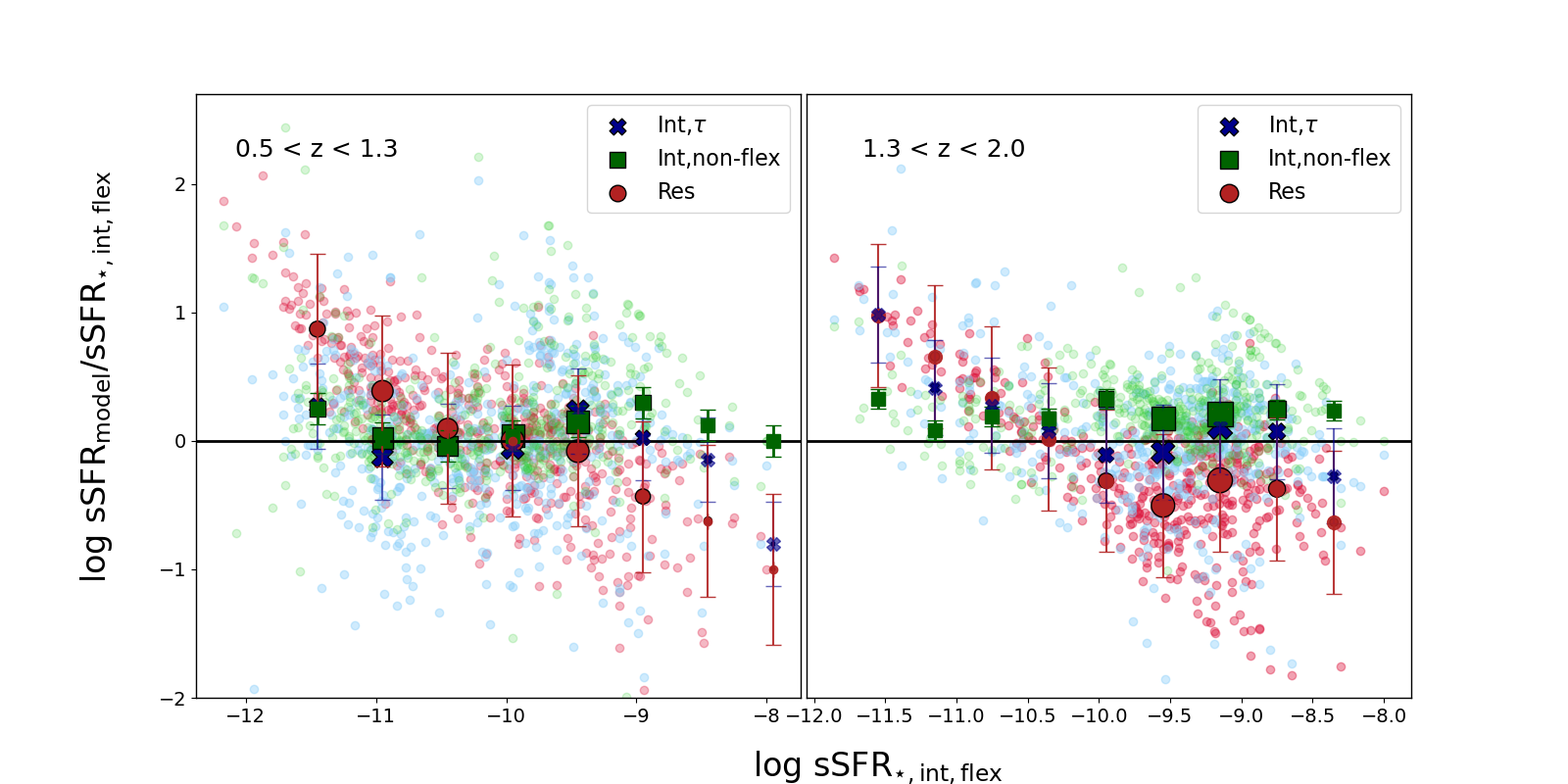}
\caption{The differences in sSFR measured over the last 100 Myr are inferred from the different model SFHs with respect to our fiducial model for galaxies with $0.5<z<1.3$ and $1.3 < z < 2.0$ (left and right panels, respectively). The figure follows the same layout as Fig.~\ref{Fig_md}. The median of average differences between the sSFR obtained from the fiducial model and the pixel-by-pixel SED fitting model are the least for the redshift range $1.3 < z < 2.0$ (For more details refer Section \ref{sSFR_results} and Figure \ref{Summary}). However, for the redshift range $0.5<z<1.3$ the median of average differences are the least for the sSFR from the parametric SFH on integrated scales from iSEDfit fitting (Green boxes) followed by the sSFR from spatially resolved SFH (red circles). The sSFR differences are the largest obtained from the fiducial model and the \texttt{Prospector} model with tau-SFH (blue crosses).}
\label{sSFR_d}
\end{figure*}

We adopt the integral of the SFH as the stellar mass throughout this work. We use the stellar masses obtained from the \texttt{Prospector} assuming a flexible non-parametric SFH (M$_{\star,\mathrm{int},\mathrm{flex}}$) to compare with the masses we get from the spatially resolved SFHs (M$_{\star,\mathrm{res}}$), with SFHs obtained by fitting the integrated photometry using a simple tau model within iSEDfit (M$_{\star,\mathrm{int},\tau}$), and the \texttt{Prospector} assuming a non-flexible SFH (M$_{\star,\mathrm{int},\mathrm{non-flex}}$). This allows us to understand the systematic differences in estimating stellar masses obtained from the different SFHs. The mass differences of M$_{\star,\mathrm{int},\tau}$, M$_{\star,\mathrm{int},\mathrm{non-flex}}$ and M$_{\star,\mathrm{res}}$ with respect to the fiducial mass (M$_{\star,\mathrm{int},\mathrm{flex}}$) is shown in \autoref{Fig_md}. 

To gain insights into the physical processes driving the observed differences over redshifts, we divided the galaxies into $0.5<z<1.3$ and $1.3<z<2.0$. For the redshift range of $0.5<z<1.3 $ (left panel), the median differences between M$_{\star,\mathrm{int},\mathrm{flex}}$ and M$_{\star,\mathrm{res}}$ is $-0.08$ dex. The absolute value of differences is nearly the same, $\sim 0.07$ dex for both redshift ranges, but it tends to be positive for the redshift range of $1.3<z<2$ (right panel). However, the median differences between M$_{\star,\mathrm{int},\mathrm{flex}}$ and M$_{\star,\mathrm{int},\tau}$ is $-0.12$ dex and $-0.14$ dex for the left and right panels respectively. Also, the median differences between M$_{\star,\mathrm{int},\mathrm{flex}}$ and M$_{\star,\mathrm{int},\mathrm{non-flex}}$ is $-0.3$ dex and $-0.36$ dex for the left and right panels respectively. 

This suggests M$_{\star,\mathrm{res}}$ are in better agreement with the masses obtained from our fiducial model M$_{\star,\mathrm{int},\mathrm{flex}}$ than M$_{\star,\mathrm{int},\tau}$ and M$_{\star,\mathrm{int},\mathrm{non-flex}}$. 
One cause for the differences in the stellar masses is the inability of  SFH$_{\star,\mathrm{int},\tau}$ and SFH$_{\star,\mathrm{int},\mathrm{non-flex}}$ to capture early, prolonged SF activity as discussed in Section \ref{Results_SFH}.

\subsection{Mass-weighted ages recovery} \label{T50_results}
To further investigate the impact of the captured details of the SF activity in the shape of the SFHs, this following section shows the comparison between mass-weighted stellar ages (t$_{50}$) from different SFHs. The t$_{50}$ of a galaxy corresponds to the age when it had assembled half of its total stellar mass. \autoref{T50_d} plots the differences in t$_{50}$ from different SFHs as a function of the t$_{50}$ obtained from our fiducial model. The differences refer to the difference between the t$_{50}$ obtained from different models (t$_{50,\star,\mathrm{res}}$, t$_{50,\star,\mathrm{int},\tau}$, t$_{50,\star,\mathrm{int},\mathrm{non-flex}}$) and the t$_{50}$ obtained from the fiducial model (t$_{50,\star,\mathrm{int},\mathrm{flex}}$). 

For the redshift range $0.5 < z < 1.3$ (left panel), the t$_{50,\star,\mathrm{res}}$ underestimates the t$_{50}$ for the galaxies by -0.14 dex  when compared to the t$_{50,\star,\mathrm{int},\mathrm{flex}}$. However, for the redshift range $1.3 < z < 2.0$ (right panel), overall the differences between the t$_{50,\star,\mathrm{res}}$ and t$_{50,\star,\mathrm{flex}}$ is the least (-0.04 dex) amongst the three described models. 

The average of median differences between the
t$_{50,\star,\mathrm{int},\tau}$ and t$_{50,\star,\mathrm{int},\mathrm{flex}}$ is -0.49 dex and -0.34 dex for left and right panels respectively. Also, the average of median differences between the
t$_{50,\star,\mathrm{int},\mathrm{non-flex}}$ and t$_{50,\star,\mathrm{int},\mathrm{flex}}$ are -0.3 dex and -0.32 dex for left and right panels respectively.  

For the galaxies at $0.5 < z < 1.3$, the SFH shape favours the younger stellar population. It is due to the formation of massive stars population in actively SF galaxies at lower redshifts, which, in turn, outshines the older stellar population leading to the skewness of the SFH towards the late cosmic time. We will discuss this in detail in Section \ref{Dis}.

To further look into the impact of outshining on the derived SFHs and the inferred galaxy properties, the following section compares the sSFR of the galaxies with the other galaxy properties. 

\subsection{Correlating sSFRs with other galaxy properties}
\label{sSFR_results}
\autoref{sSFR_d} plots the differences in sSFR measured over the last 100 Myr obtained from different SFHs as a function of the sSFRs obtained from our fiducial model. The differences refer to the difference between the sSFR obtained from the three models (sSFR$_{\star,\mathrm{res}}$, sSFR$_{\star,\mathrm{int},\tau}$, sSFR$_{\star,\mathrm{int},\mathrm{non-flex}}$) and the sSFR obtained from the fiducial model (sSFR$_{\star,\mathrm{int},\mathrm{flex}}$). 

The left and right panel plots the sSFRs in the two redshift ranges: $0.5 < z < 1.3$ and $1.3 < z < 2.0$. For galaxies with low sSFR values (log (sSFR$_{\star,\mathrm{int},\mathrm{flex}}/\mathrm{yr}) < -10$) inferred from the fiducial model, the sSFR$_{\star,\mathrm{res}}$ tends to provide somewhat higher estimates compared to those inferred from the fiducial model.
In contrast, for galaxies with high sSFR values (log (sSFR$_{\star,\mathrm{int},\mathrm{flex}}/\mathrm{yr}) > -10$) from the fiducial model, sSFR$_{\star,\mathrm{res}}$ either slightly underestimates or traces the sSFR$_{\star,\mathrm{int},\mathrm{flex}}$ quite well. The average median differences between sSFR$_{\star,\mathrm{res}}$ and sSFR$_{\star,\mathrm{int},\mathrm{flex}}$ are -0.1 dex for $0.5 < z < 1.3$ and -0.02 dex for $1.3 < z < 2.0$. This overestimation of sSFR for galaxies with little or no SF and the underestimation of sSFR for the actively star-forming galaxies when compared to sSFR$_{\star,\mathrm{int},\mathrm{flex}}$ can be attributed to the SFH model choice for each of the pixels. We will discuss this later in detail in Section \ref{Dis}.
On the other hand, the average median differences between sSFR$_{\star,\mathrm{int},\mathrm{non-flex}}$ and sSFR$_{\star,\mathrm{int},\mathrm{flex}}$ are 0.1 dex and 0.22 dex for low and high redshift galaxies, respectively. Furthermore, the sSFR$_{\star,\mathrm{int},\tau}$ shows a different trend, where the average median differences between sSFR$_{\star,\mathrm{int},\tau}$ and sSFR$_{\star,\mathrm{int},\mathrm{flex}}$ are $-$0.08 dex for low redshift galaxies (left panel) and 0.16 dex for high redshift galaxies (right panel). 

Overall, the average median differences in our results show that the galaxy properties inferred from the SFH$_{\star,\mathrm{res}}$ are in better agreement with those inferred with the flexible \texttt{Prospector} SFH model (the fiducial model) when compared to the properties inferred from the other two approaches. 

\begin{figure*} 
\includegraphics[width=\textwidth]{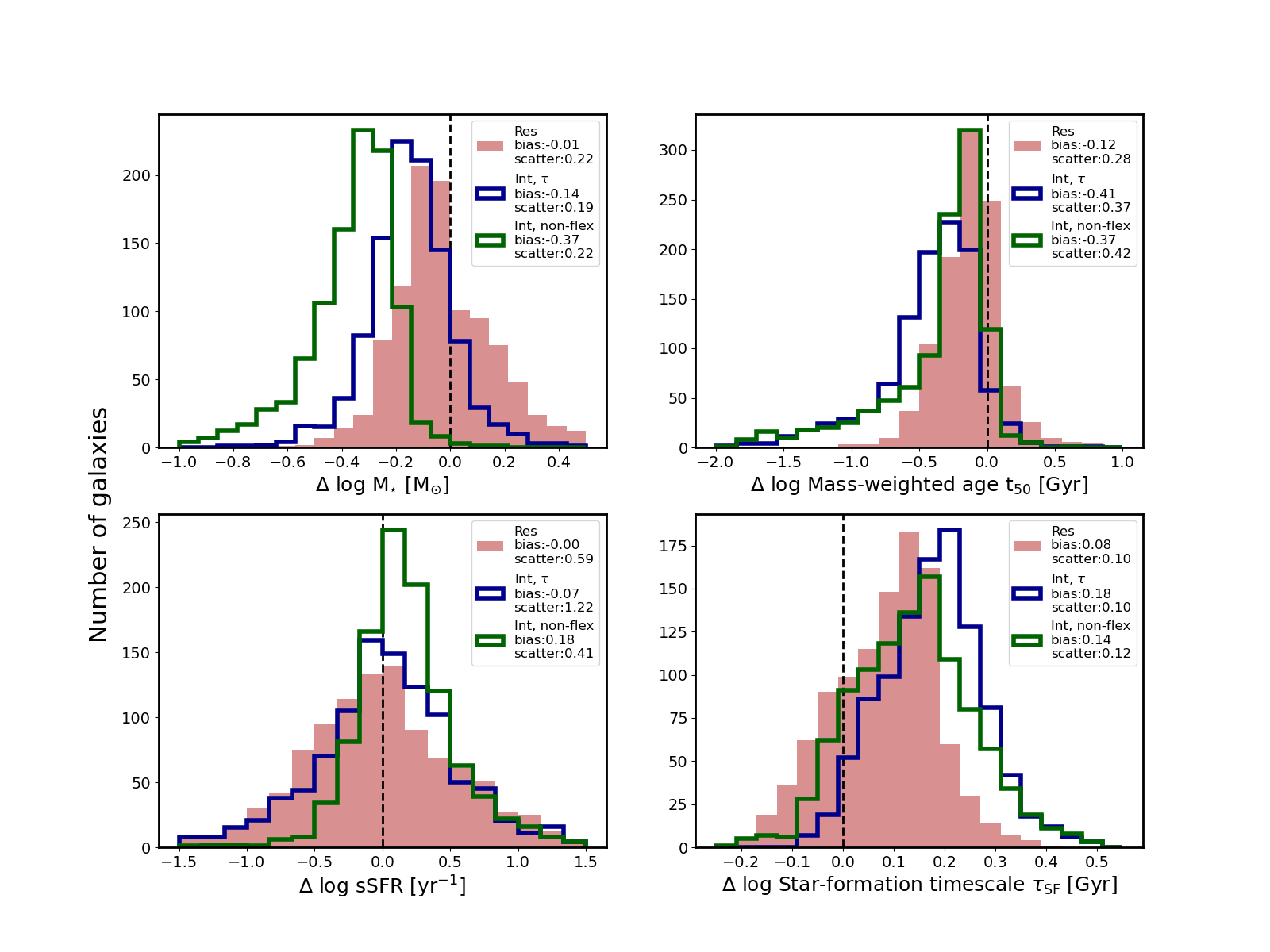}
\caption{Distribution of the inferred galaxy properties from the model SFHs offsets from the properties inferred using the fiducial model (\texttt{Prospector} model with a flexible SFH). The models include the spatially resolved model (red histograms), a simple tau model within iSEDfit (blue), and \texttt{Prospector} model that assumes tau-SFH (green). The fiducial model is the \texttt{Prospector} model that adopts flexible SFH (dashed black line). The offsets in stellar mass (top left), mass-weighted age (top right), sSFR (bottom left) and star-formation timescale (bottom right) are shown. The measured bias (average offset) and scatter (standard deviation) are indicated in the legends of the panels. The properties inferred from the spatially resolved SFHs are in best agreement with the properties inferred from the fiducial, flexible SFHs.}
\label{Summary}
\end{figure*}

\section{DISCUSSION} \label{Dis}

In this section, we discuss the implications of how the assumed SFH models and the spatial resolution affect the inferred physical properties of the galaxies. Overall, our analysis supports that the SFHs obtained from spatially resolved scales (SFH$_{\star,\mathrm{res}}$) provide insights into the galaxies' internal structure and assembly history better than the SFHs obtained from simple parametric forms on the integrated scales (SFH$_{\rm \star, int,\tau}$ and SFH$_{\rm \star, int,non-flex}$). 
We will conclude this section by discussing the limitations of this work.

\subsection{The need of flexibility in SFH} \label{Need_Flex}
Section \ref{Results_SFH} and Figure \ref{SFHs} illustrate a strong correspondence between the SFH$_{\star,\mathrm{res}}$ and the flexible SFH model of \texttt{Prospector} (SFH$_{\rm \star, int,flex}$; our fiducial SFHs) using integrated photometry. Additionally, it points out the challenge of SFH$_{\rm \star, int,\tau}$ and SFH$_{\rm \star, int,non-flex}$ to match the amplitude of the fiducial SFH$_{\rm \star, int,flex}$, especially at large lookback times for most galaxies. 

The pixel-by-pixel SED fitting model allows for variations in the best-fit values of stellar mass, stellar age, and timescale ($\tau$) for each pixel, resulting in spatially resolved colour gradient; mass,  age, and $\tau$ maps, as shown in Figure \ref{F_maps} (Broader Parameter Space). This, in turn, leads to a more stochastic SFH$_{\star,\mathrm{res}}$ compared to SFHs constructed using a single best-fit value of stellar mass, stellar age and $\tau$ from the integrated scales (SFH$_{\rm \star, int,\tau}$ and SFH$_{\rm \star, int,non-flex}$). The burstiness in SF represents the ISM physics and feedback processes, acting on spatial and temporal scales within galaxies, that outline the galaxy's evolutionary pathways \citep{Kauffmann_2006,Tacchella_2016,Semenov_2017,Semenov_2018,Iyer_2020,Semenov_2021}. 

Figure \ref{SFH_resolved} demonstrates how SFH$_{\star,\mathrm{res}}$ accurately captures the complex SFH with burstiness in SF activity occurring on kpc scales rather than galaxy-wide, represented by the peaks and lows of SFR. Furthermore, the observed stochasticity of the SFHs reveals the older population of stars that otherwise remains hidden due to the outshining effects \citep{Sawicki_1998,Papovich_2001,2010MNRAS.407..830M, Pforr_2012}. 
Our findings agree with the work of \citet{https://doi.org/10.48550/arxiv.2212.08670}, which demonstrated that a spatially resolved analysis could reveal the existence of older underlying stellar populations that are otherwise outshined in integrated analyses, significantly impacting our understanding of these galaxies' nature \citep[e.g.][]{Zibetti_2009,Pforr_2012,Sorba_2015}. This is because the domination of strong emission lines drives the need to fit the integrated light with extremely young stellar populations. This explains the shift of the peaks in SFH$_{\rm \star, int,\tau}$ and SFH$_{\rm \star, int,non-flex}$ towards the late cosmic times and the extension of SFH$_{\star,\mathrm{res}}$ over a wider age range of galaxies (see Figure \ref{SFHs}).
   
Therefore, SFH$_{\star,\mathrm{res}}$ can better trace the fiducial SFH$_{\rm \star, int,flex}$ than SFH$_{\rm \star, int,\tau}$ and SFH$_{\rm \star, int,non-flex}$.
Moreover, it can provide a deeper understanding of the physical mechanisms that govern SF within galaxies. Besides shedding light on the SF activity within galaxies, we aim to investigate whether spatially resolved observations can enable us to infer galaxies' properties consistently, when compared to inferred properties from the SFH$_{\rm \star, int,\tau}$ and SFH$_{\rm \star, int,non-flex}$. For this, the following section explores the reasons for the consistency and biases observed in the galaxy properties inferred from different model SFHs presented in Section \ref{Results}.

\subsection{Impact of assumed spatially resolved SFHs on galaxy properties}

In Section \ref{Results}, we presented a systematic discrepancy in inferred galaxy properties from different SFH models when compared to the fiducial SFH$_{\rm \star, int, flex}$. 
However, this discrepancy is within the uncertainties of galaxy properties' estimates due to stellar synthesis modelling \citep{Conroy_2009}.
In this section, we will try to understand the reason for this discrepancy in the inferred properties of the galaxies from different SFH approaches. \autoref{Summary} summarises these biases, which shows that the SFH$_{\star,\mathrm{res}}$ have the least and closest to 0 offsets for stellar mass, t$_{50}$, sSFR and $\tau_{\mathrm{SF}}$ from those inferred using fiducial SFH$_{\rm \star, int,flex}$.

\begin{figure*}
\centering
\includegraphics[width=7.5in]{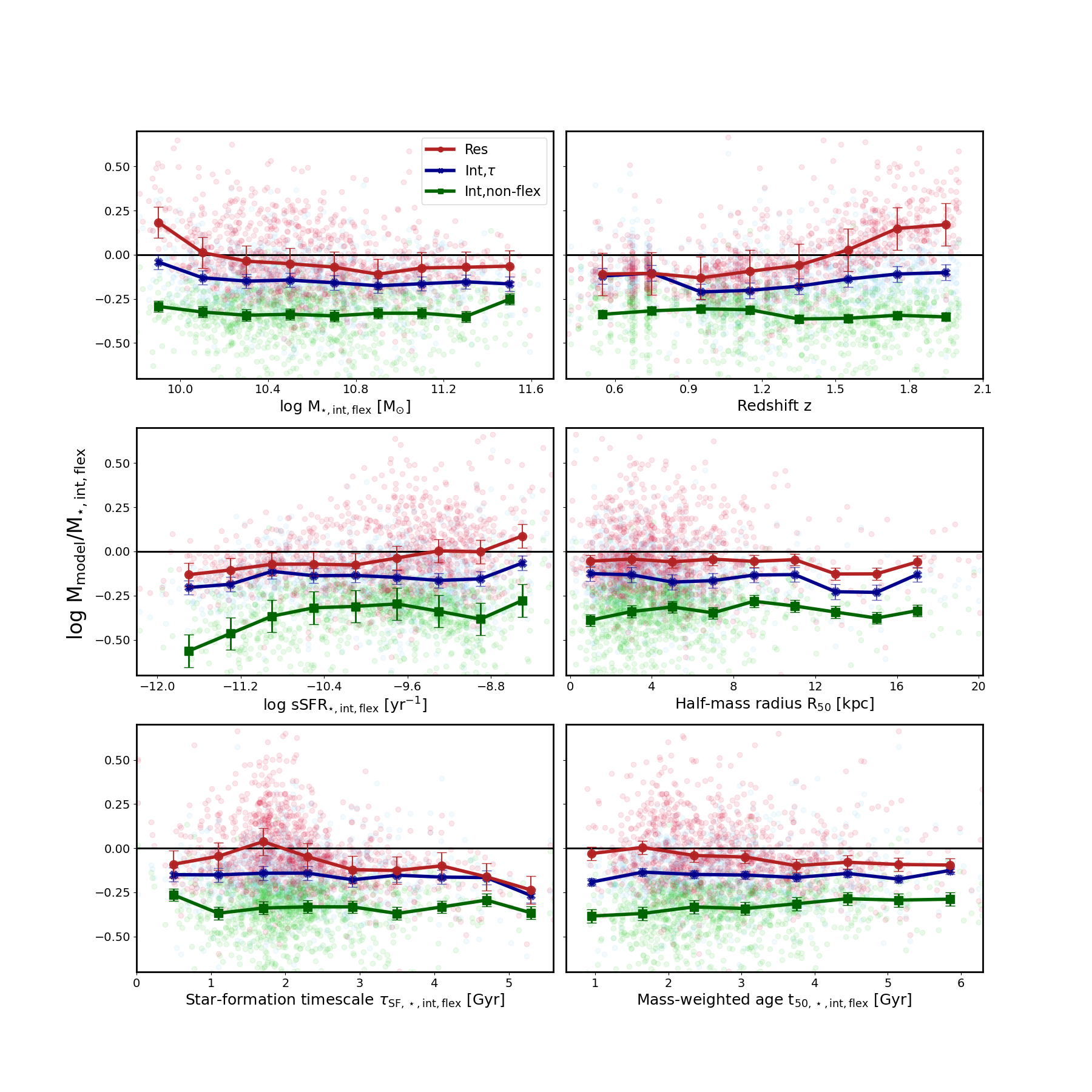}
\caption{Differences in the stellar masses obtained from the different model SFHs with respect to galaxy properties derived from our fiducial model as a function of stellar mass (top left), redshift (top right), sSFR (middle left), half-mass radius (middle right), star-formation timescales (bottom left) and mass-weighted ages (bottom right). The models include the spatially resolved model (red circles),  a simple tau model within iSEDfit (blue crosses), and \texttt{Prospector} model that assumes tau-model (green boxes). The fiducial model is the \texttt{Prospector} model that adopted flexible SFH. The mass differences obtained from the fiducial model and the pixel-by-pixel SED fitting model are the least, followed by the differences obtained from the fiducial and the tau model within iSEDfit. The mass differences are the largest obtained from the fiducial model and the \texttt{Prospector} model with tau-SFH. Through the analysis of mass differences with sSFR, t$_{50}$, and $z$, a notable trend emerges: the agreement between masses obtained from spatially resolved SFHs and fiducial masses is considerably better for actively star-forming, young galaxies, whereas the discrepancy remains for the other two models and the fiducial model. }
\label{Fig_md_m}
\end{figure*}

\subsubsection{Recovered Stellar Masses}

Section \ref{R_Diff_mass} presented in detail the biases introduced in stellar mass estimates by different assumed SFHs on spatially resolved and unresolved scales when compared with those inferred from the fiducial SFH$_{\rm \star, int, flex}$.

To understand the bias, we plot the stellar mass differences against other physical properties to determine any trends we might be missing. Figure \ref{Fig_md_m} plots the stellar mass differences against the stellar masses, redshifts, sSFR, half-light mass radius (R$_{50}$), t$_{50}$ and $\tau_{\mathrm{SF}}$. In the study by \citet{Sorba_2018}, the observed inconsistency was attributed to the outshining effect \citep[see also][]{Papovich_2001, Maraston_2010, Conroy_2013}.

We investigate this by focusing on the stellar mass differences against the redshifts $z$, sSFR, and t$_{50}$. We observe that the discrepancy in the inferred masses from SFH$_{\star,\mathrm{res}}$ almost vanishes for actively star-forming galaxies (sSFR $> 10^{-10}$ yr$^{-1}$), i.e., the galaxies with younger stellar populations. However, the SFH$_{\rm \star, int,\tau}$ and SFH$_{\rm \star, int,non-flex}$ still underestimate the inferred stellar mass estimates for these galaxies. On the other hand, the galaxies with the lower t$_{50}$, implying the galaxies which recently assembled 50$\%$ of the stellar masses are expected to be dominated by younger stellar populations. In these younger galaxies, the observed discrepancy between the SFH$_{\star,\mathrm{res}}$ and fiducial SFH$_{\rm \star, int, flex}$ almost vanishes. However, the underestimation of masses from SFH$_{\rm \star, int,\tau}$ and SFH$_{\rm \star, int,non-flex}$ remains. This is because the young stellar population outshines the older population, resulting in an omission of a significant portion of the older stellar masses formed in the galaxies. 
Similar reasoning can be applied for the observed underestimation of masses from SFH$_{\rm \star, int,\tau}$ and SFH$_{\rm \star, int,non-flex}$ with the redshift of the galaxies. The underestimation of masses can be attributed to the dominance of young stellar populations and hence outshining effects. 
 Furthermore, we observe no noticeable trend of the mass differences with stellar masses, $\tau_{\mathrm{SF}}$ and R$_{50}$. 

Therefore, the work highlights the importance of the spatially resolved scales to recover the older stellar masses in galaxies otherwise obscured due to the outshining effects.

\subsubsection{Recovered mass-weighted ages and sSFRs}

As shown in Section \ref{T50_results}, according to our analysis, all three SFH models, including SFH$_{\star,\mathrm{res}}$ and parametric SFHs obtained from integrated scales (SFH$_{\rm \star, int,\tau}$ and SFH$_{\rm \star, int,non-flex}$), tend to provide lower estimates of t$_{50}$ when compared to the fiducial SFH$_{\rm \star, int, flex}$. It is important to note that in these comparisons, the flexible Prospector model, SFH$_{\rm \star, int, flex}$, serves as our reference rather than the ground truth. Specifically, for redshifts $0.5 < z < 1.3$, the  t$_{50}$ estimates inferred using SFH$_{\star,\mathrm{res}}$ are lower by 0.14 dex, while t$_{50}$ estimates from SFH$_{\rm \star, int,\tau}$ and SFH$_{\rm \star, int,non-flex}$ are lower by  $>0.25$ dex relative to the t$_{50}$ inferred using SFH$_{\rm \star, int, flex}$.
However, we find that for redshifts $1.3 < z < 2.0$, the differences between the t$_{50}$ values obtained from SFH$_{\star,\mathrm{res}}$ and the ones from the fiducial SFH$_{\rm \star, int, flex}$ almost vanish ($\sim$0.04 dex). On the other hand, the differences remain larger than 0.25 dex for SFH$_{\rm \star, int,\tau}$ and SFH$_{\rm \star, int,non-flex}$ with SFH$_{\rm \star, int, flex}$ \citep{Carnall_2019, Leja_2019}.

Our study suggests that the significant lower estimates of the t$_{50}$ distributions in lookback time with the t$_{50}$ estimates from SFH$_{\rm \star, int, flex}$ is due to the skewed SFHs towards the late cosmic time for SFH$_{\rm \star, int,\tau}$ and SFH$_{\rm \star, int,non-flex}$. This skewness, primarily due to outshining effects, obscures the extended process of early galaxy formation and affects the estimated t$_{50}$ by several orders of magnitude 
Furthermore, these lower estimated values of t$_{50}$ agree with the findings of \citet{Suess_2022}, who also reported extremely young inferred ages of the stellar population from the SFHs obtained on integrated scales.

Moreover, the higher estimated values of sSFR over the last 100 My by sSFR$_{\star,\mathrm{int},\tau}$ and sSFR$_{\star,\mathrm{int},\mathrm{non-flex}}$ when compared to the sSFR$_{\star,\mathrm{int},\mathrm{flex}}$ could also be due to the problem of outshining as stellar ages are intricately linked to the determination of sSFRs. As discussed above, the priors that inform the stellar age distribution can lead to the shifted peak of SFH towards the late cosmic time. The skewed peak, in turn, results in the overestimation of sSFR (See Figure \ref{sSFR_d}). However, from SFH$_{\star,\mathrm{res}}$, we observe the sSFRs over the last 100 My are underestimated for the actively star-forming galaxies (log (sSFR$_{\star,\mathrm{int},\mathrm{flex}}) > -10$) according to the sSFR$_{\star,\mathrm{int},\mathrm{flex}}$. This could be because of the trade-off between accurately inferring stellar age and sSFR for a declining SFH defined by the tau model. When adding the SFHs of all the pixels, the incorrect position of the SFHs' peak, determined by the $\tau$ parameter of the model, prevents the recovery of the correct sSFR for recent times.
The simple tau model adopted for defining the SFH of each pixel does not consider the increasing SFR. This could be one reason for the inaccuracy of the SFH$_{\star,\mathrm{res}}$ to infer the accurate sSFRs when compared to the fiducial sSFR$_{\star,\mathrm{int},\mathrm{flex}}$ for the highly star-forming galaxies.

On the other hand, for lower estimates of fiducial sSFR$_{\star,\mathrm{int},\mathrm{flex}}$, our analysis reveals that even for galaxies with minimal or no SF, SFH$_{\star,\mathrm{res}}$ still infers a significant SFR. This is due to the uncertainties associated with the stellar ages of a few of the pixels. The stellar age is one parameter that defines the SFR of the pixels, hence, contributing to the overall SFH of the galaxy. These uncertainties in stellar ages can be attributed to the overestimation of sSFR for galaxies with little or no SF.

To address these issues, incorporating other SFH parametrisations, such as the delayed tau model, to define the SFH of each pixel can be tested. This would allow for a rising SFH for both early and late cosmic time for each pixel's SFH. However, testing these parametrisations is outside the scope of this paper. The key takeaway here is that for actively star-forming galaxies (those with log (sSFR$_{\star,\mathrm{int},\mathrm{flex}}) > -10$), sSFR inferred from SFH$_{\star,\mathrm{res}}$ are not overestimated when compared to those obtained using the fiducial SFH$_{\star,\mathrm{int},\mathrm{flex}}$, indicating that the SFHs obtained from the pixel-by-pixel SED fitting method could counteract the outshining effects.

In summary, spatially resolved SFHs offer a more effective approach to counteract the outshining effects when determining the physical properties of galaxies. 

\subsection{Limitations and future outlook} 
\label{Lim}
Although this work's pixel-by-pixel SED fitting approach demonstrates the consistency of inferred galaxy properties, it is only the first step in gaining new insights into the galaxy evolution and formation process. We can use the inferred galaxy properties to estimate and compare the growth in the center versus outskirts of galaxies. This can further shed light on inside-out growth pattern of galaxies. Here are a few considerations that should be kept in mind for additional future work:

\begin{itemize}
    \item Different assumptions such as additional random burst on top of a constant or delayed SFH can alter the estimated stellar masses and other galaxy properties \citep{Mosleh_2020, Carnall_2019}. Incorporating these other SFH parametrisations to define the SFH of each pixel can be tested to better constrain the inferred physical properties of the galaxies.
    \item When analyzing data with a certain pixel scale, the best level of detail is achieved by examining the resolved SEDs within each pixel. However, it is important to consider the signal-to-noise within these pixels, particularly towards the outskirts. In some cases, the signal-to-noise may be too low, leading to unreliable estimations of resolved stellar population properties. To address this issue, we can apply the Voronoi binning method \citep{Cappellari_2003} that groups pixels based on reaching a desired S/N threshold in multiple resolved filters.
    \item Additionally, it is worth noting that the signal in the data exhibit a correlation between adjacent pixels. This effect is particularly important when the spatial resolution of the instrument, represented by the point spread function (PSF), is larger than the size of the individual pixels. In such cases, neighboring pixels may share some level of information, potentially impacting the accuracy of derived galaxy properties. 
    \item Having an absolute truth against which we could compare our derived quantities would be ideal. A possible approach is to conduct our analysis of pixel-by-pixel SED fitting on mock observations from 3D radiative transfer calculations from hydrodynamical simulation \citep[e.g.,][]{2018MNRAS.476.1705S, Lower_2020,2022ApJ...930...66Q,2022MNRAS.513.2904T}. 
\end{itemize}

\section{Conclusions} \label{Conc}

We present detailed measurements of SFHs both on global and spatially-resolved scales for a sample of $\sim$970 distant galaxies with redshifts $z = 0.5-2.0$ to better understand the systematics involved when estimating galaxy properties. On spatially resolved scales, we derive the SFH of a galaxy by summing the SFHs of individual pixels obtained using pixel-by-pixel SED fitting adopting iSEDfit (SFH$_{\rm \star, res}$). On global scales, we fit the integrated photometry using (i) a simple tau model within iSEDfit (SFH$_{\rm \star, int,\tau}$), (ii) a simple tau model within \texttt{Prospector} (SFH$_{\rm \star, int,non-flex}$), and (iii) a flexible, non-parametric model within \texttt{Prospector} (SFH$_{\rm \star, int,flex}$), which we adopted as our fiducial model for the comparison.

Our main findings and conclusions are following:

\begin{itemize}    
    \item Both SFH$_{\rm \star, res}$ from spatially resolved scales and SFH$_{\rm \star, int,flex}$ from the flexible \texttt{Prospector} model lead to a large diversity of inferred SFHs. Importantly, as shown in Fig. \ref{SFHs} (see also Fig. \ref{SFH_resolved}), SFH$_{\rm \star, res}$ and SFH$_{\rm \star, int,flex}$ agree well with each other, while more simplistic, tau-based models (SFH$_{\rm \star, int,\tau}$ and SFH$_{\rm \star, int,non-flex}$) are not able to capture this large diversity: they are only consistent with SFH$_{\rm \star, res}$ and SFH$_{\rm \star, int,flex}$ in recent lookback times, while missing early star formation. 
    \item This has direct consequences on the inferred stellar population parameters, in particular the stellar masses (Fig. \ref{Fig_md}), stellar ages (Fig. \ref{T50_d}) and sSFR (Fig. \ref{sSFR_d}). Specifically, we find a median stellar mass difference of $\sim$ 0.1-0.4 dex ($\sim 25 \% -  152 \%$) between the masses obtained from unresolved, tau-model SFHs (SFH$_{\rm \star, int,\tau}$ and SFH$_{\rm \star, int,non-flex}$) and the fiducial SFH$_{\rm \star, int,flex}$, which reduces to only $\sim$ 0.07 dex ($\sim 18 \%$) when using SFH$_{\rm \star, res}$ and SFH$_{\rm \star, int,flex}$. Similarly, mass-weighted ages are lower by 0.3-0.5 dex ($\sim 99 \% - 217 \%$) in the case of SFH$_{\rm \star, int,\tau}$ and SFH$_{\rm \star, int,non-flex}$ in comparison with SFH$_{\rm \star, int,flex}$, while this difference reduces significantly (to 0.1 dex; $\sim 25 \%$) when comparing the ages from SFH$_{\rm \star, res}$ and SFH$_{\rm \star, int,flex}$. 
    \item These differences are connected: the limited flexibility of the SFH shape of the tau model captures only the recent SFH, thereby missing early star formation and hence underestimates the stellar age and stellar mass. The tau-model mainly captures the recent SFH because the young stellar populations dominate the SED, meaning that the younger stellar populations are outshining the older stellar populations. Both the SFH$_{\rm \star, res}$ from spatially resolved scales and the SFH$_{\rm \star, int,flex}$ from the flexible \texttt{Prospector} model are less affected from outshining because outshining typically only affects certain spatial regions and the prior in the non-parametric SFH approach weights against very young stellar populations, respectively. 
\end{itemize}

In summary, the SFHs on spatially resolved scales motivate flexible SFHs on global scales. In light of JWST and high-redshift galaxies, in which SFHs are bursty and outshining is a crucial factor, detailed studies of the SFH on spatially resolved scales \textit{in connection} with flexible SFHs on global scales are needed in the future.

\section*{Data Availability}

Data available on request.



\bibliographystyle{mnras}
\bibliography{example} 


\appendix
\section{Recovering Star-formation timescales}
\begin{figure*}
\includegraphics[width=\textwidth]{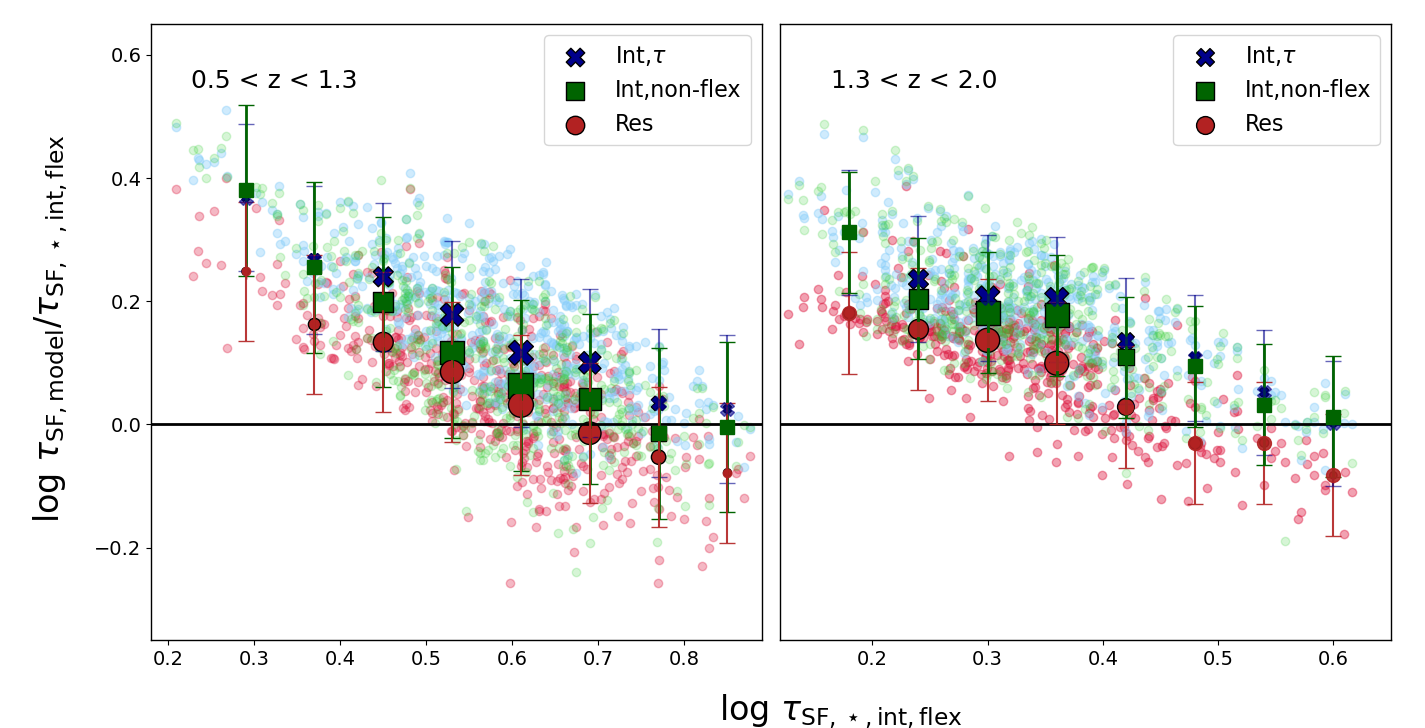}
\caption{The differences in star-formation timescale obtained from the different model SFHs with respect to our fiducial model for objects with $0.5 < z < 1.3$ and $1.3 < z < 2.0$ (left and right panel respectively). The different models include the spatially resolved model (red circles), a simple tau model within iSEDfit (blue crosses) and \texttt{Prospector} model which assumes tau-SFH (green boxes). The fiducial model is the \texttt{Prospector} model which assumes flexible SFH.}
\label{T28}
\end{figure*}

\autoref{T28} plots the differences in $\tau_{\mathrm{SF}}$ from different SFHs as a function of the $\tau_{\mathrm{SF}}$ obtained from our fiducial model. The differences refer to the difference between the $\tau_{\mathrm{SF}}$ obtained from different models ($\tau_{\mathrm{SF},\star,\mathrm{res}}$, $\tau_{\mathrm{SF},\star,\mathrm{int},\tau}$, $\tau_{\mathrm{SF},\star,\mathrm{int},\mathrm{non-flex}}$) and the $\tau_{\mathrm{SF}}$ obtained from the fiducial model ($\tau_{\mathrm{SF},\star,\mathrm{int},\mathrm{flex}}$). 

For the redshift range $0.5 < z < 1.3$ (left panel), the $\tau_{\mathrm{SF},\star,\mathrm{res}}$ on an average overestimates the $\tau_{\mathrm{SF}}$ for the galaxies when compared to the $\tau_{\mathrm{SF},\star,\mathrm{int},\mathrm{flex}}$. However, for both the redshift ranges $0.5 < z < 1.3$ (left panel) and $1.3 < z < 2.0$ (right panel), overall the differences between the $\tau_{\mathrm{SF},\star,\mathrm{res}}$ and $\tau_{\mathrm{SF},\star,\mathrm{int},\mathrm{flex}}$ is the least among the three described models. The average of median differences between $\tau_{\mathrm{SF},\star,\mathrm{res}}$ and $\tau_{\mathrm{SF},\star,\mathrm{int},\mathrm{flex}}$ is $\sim$ 0.06 dex for both the redshift ranges. However, the average median differences between the
$\tau_{\mathrm{SF},\star,\mathrm{int},\tau}$ and $\tau_{\mathrm{SF},\star,\mathrm{int},\mathrm{flex}}$ is 0.17 dex and 0.16 dex for left and right panels respectively. Also, the average median differences between the
$\tau_{\mathrm{SF},\star,\mathrm{int},\mathrm{non-flex}}$ and $\tau_{\mathrm{SF},\star,\mathrm{int},\mathrm{flex}}$ are 0.13 dex and 0.14 dex for left and right panels respectively.  

The $\tau_{\mathrm{SF}}$ is a crucial parameter for understanding how galaxies evolve through SF activity. However, accurately determining this timescale from SFHs obtained using integrated scales can be challenging due to the limited prior parameter space. The estimated $\tau_{\mathrm{SF}}$ by the simple parametric SFHs on the integrated scales may only be able to account for a few of the peaks associated with different SF phenomena. As a result, these SFHs can either underestimate or overestimate the timescale.

In particular, in our case, the overestimation of the $\tau_{\mathrm{SF}}$ by $\tau_{\mathrm{SF},\star,\mathrm{int},\tau}$ and $\tau_{\mathrm{SF},\star,\mathrm{int},\mathrm{non-flex}}$ when compared to $\tau_{\mathrm{SF},\star,\mathrm{int},\mathrm{flex}}$ occur due to the outshining effect. This effect causes most of the SF to occur later in cosmic time, which can skew the SFH towards later times and lead to an overestimation of $\tau_{\mathrm{SF}}$. However, when we consider $\tau_{\mathrm{SF},\star,\mathrm{res}}$, this overestimation almost disappears, reducing to an average of 0.06 dex.
 
Therefore, it is crucial to consider the limitations of simple parametric SFHs when estimating the $\tau_{\mathrm{SF}}$. Additionally, the results highlight that the SFH on spatially resolved scales can better recover the $\tau_{\mathrm{SF}}$ and hence, we can better understand the timescale at which galaxy evolution occurs through SF activity.

\section{Reliability of the Proposed Flexibility} 

\begin{figure*}
\includegraphics[width=\textwidth]{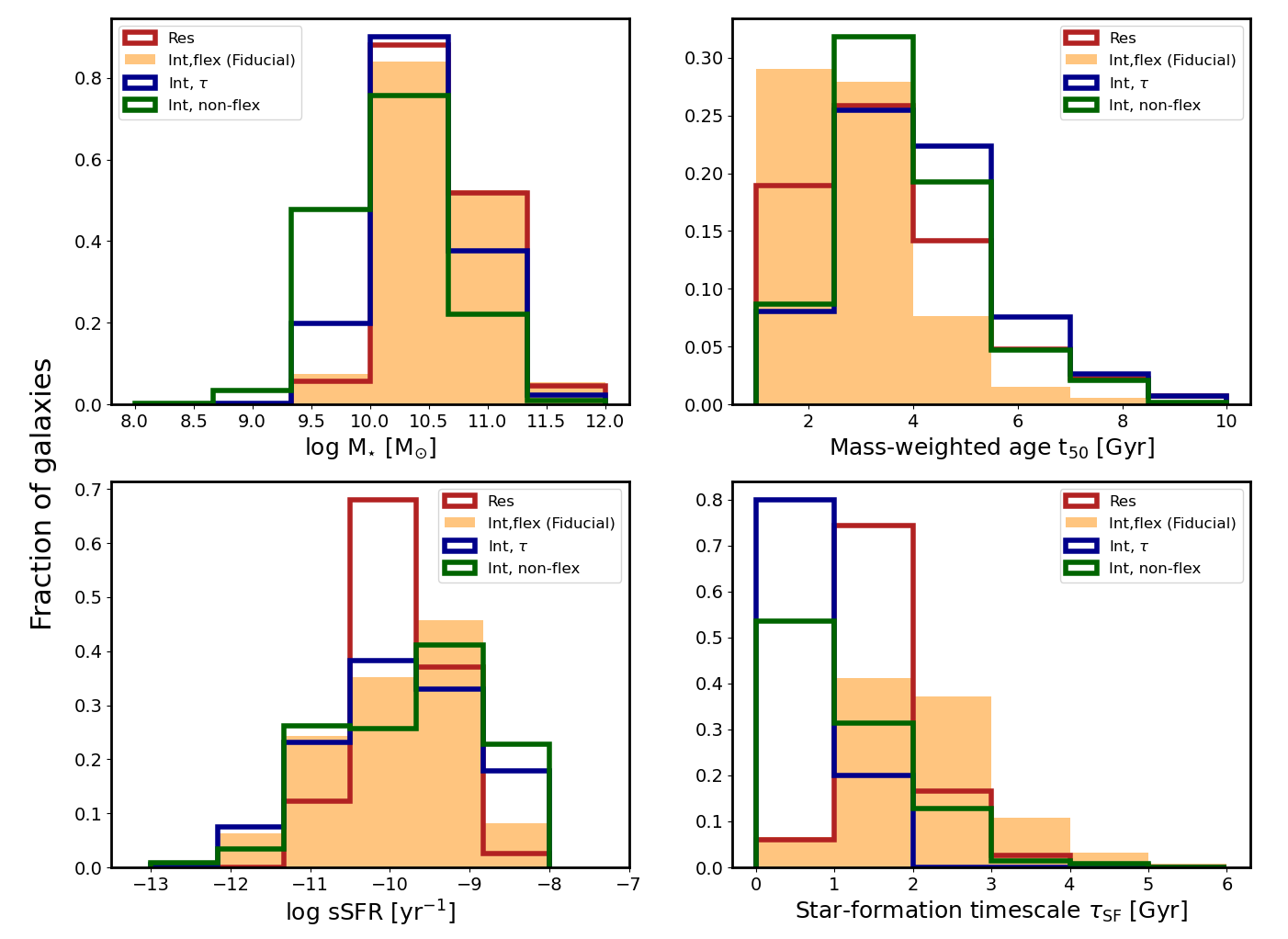}
\caption{The plot shows the distribution of inferred galaxy properties from different model SFHs and the fiducial one. The models include the spatially resolved model (red),  a simple tau model within iSEDfit (blue), and \texttt{Prospector} model that assumes tau-SFH (green). The fiducial model is the \texttt{Prospector} model that adopts flexible SFH (orange histograms). Top left panel: Inferred stellar masses of galaxies. Top right panel: Mass-weighted age distribution. Lower left panel: sSFR distribution. Lower right panel: Star formation timescale distribution. }
\label{Hist_perc}
\end{figure*}

In Figure \ref{Hist_perc}, we present the distribution of inferred galaxy properties from different model SFHs in our study. The solid orange bars show the distribution of the fraction of galaxies for each bin/range of the galaxy properties inferred from the fiducial model. The galaxy properties from the other three models we compare include the spatially resolved model (red), a simple tau model within iSEDfit (blue), and \texttt{Prospector} model that assumes tau-SFH (green).

We find that the physical properties inferred form SFH$_{\star,\mathrm{res}}$ exhibits the highest level of consistency with the fiducial SFH$_{\star,\mathrm{int},\mathrm{flex}}$.
This is evident as the majority of galaxy properties inferred from SFH$_{\star,\mathrm{res}}$ trace the stellar mass, t$_{50}$, and $\tau_{\mathrm{SF}}$ bins of those inferred using fiducial SFH$_{\star,\mathrm{int},\mathrm{flex}}$, with only $\sim 2\%$, $\sim 17\%$, and $\sim 33\%$ galaxies falling outside these boundaries. In contrast, when considering SFH$_{\star,\mathrm{int},\tau}$, a larger proportion of galaxies ($\sim 12\%$, $\sim 34\%$, and $\sim 72\%$) are unable to trace the stellar mass, t$_{50}$, and $\tau_{\mathrm{SF}}$ bins inferred using fiducial SFH$_{\star,\mathrm{int},\mathrm{flex}}$.
Similarly, SFH$_{\star,\mathrm{int},\mathrm{non-flex}}$ results in $\sim 28\%$, $\sim 30\%$, and $\sim 46\%$ of galaxies falling outside these bins. For sSFR, the percentage of galaxies falling outside of the galaxy properties' bins from the fiducial model are $\sim 26\%$, $\sim 11\%$, and $\sim 11\%$ for the three models, respectively. As discussed in Section \ref{Dis}, this can be attributed to the trade-off between the estimation of mass-weighted age and sSFR.

\bsp	
\label{lastpage}
\end{document}